\documentclass[cits]{PoS}

\newcommand{\be}{\begin{equation}}
\newcommand{\ee}{\end{equation}}
\newcommand{\bea}{\begin{eqnarray}}
\newcommand{\eea}{\end{eqnarray}}

\newcommand{\msbar}{\overline{\mathrm{MS}}}

\newlength{\MiniPageLeft}
\newlength{\MiniPageRight}

\title{Light dynamical fermions on the lattice:\\ 
       toward the chiral regime of QCD\footnote{CERN-PH-TH/2006-221}}

\ShortTitle{Light dynamical fermions}

\author{\speaker{Leonardo Giusti}\thanks{On leave from Centre de Physique Th\'eorique, 
        CNRS Luminy, F-13288 Marseille, France}\\
        CERN, Physics Department, TH Division, CH-1211
        Geneva 23, Switzerland\\
        E-mail: \email{Leonardo.Giusti@cern.ch}}

\abstract{Algorithmic and technical progress achieved over the 
last few years makes QCD simulations with light dynamical quarks much faster 
than before. As a result lattices with pions as light as 250--300 MeV can be 
simulated with the present generation of computers. I review recent 
conceptual and numerical progress in this field, with particular emphasis on results obtained 
and difficulties encountered in simulations with significantly smaller quark 
masses with respect to previous computations. I also attempt to compare physical 
results for pion masses and decay constants available to date in the two-flavour 
theory with expectations from chiral perturbation theory.
}

\FullConference{XXIVth International Symposium on Lattice Field Theory\\
		 $23$--$28$ July 2006\\
		 Tucson, Arizona, USA}

\begin{document}

\section{Introduction}
Lattice field theory provides the only known non-perturbative 
regularization of QCD where computations can be carried out 
from first principles. At finite volume and lattice spacing the 
Euclidean functional integrals can be computed non-perturbatively 
by numerical simulations. All the systematics associated with those 
calculations can, at least in principle, be quantified and eventually 
removed by exploiting the properties of the underlying quantum field 
theory, better numerical algorithms, faster computers, etc., but
without adding extra free parameters or dynamical assumptions 
in the theory. For more than twenty years the available computer power 
confined lattice QCD to the so-called quenched approximation, where the 
fermion determinant in the effective gluon action is replaced by its average 
value. Even though the discrepancy of quenched results with experimental 
data is moderate for several simple physical observables, see for example 
Ref.~\cite{Aoki:2002fd}, quenched QCD is not a systematic approximation 
of the theory and estimates of the corresponding errors are not reliable. 
Full QCD simulations are needed for first-principle results. 
Moreover many interesting processes where quark-antiquark pair 
production and/or unitarity play a crucial r\^ole, such as 
$\rho\rightarrow\pi\pi$ decays, $\eta$--$\eta'$ splitting, 
neutron electric dipole moment, etc., can only be addressed
in simulations with dynamical quarks. 

The first large-scale full QCD simulations with interesting 
lattice spacings and volumes were started in the second part of the 
90s \cite{Eicker:1998sy,Lippert:1997vg,Allton:2001sk,AliKhan:2001tx,Aoki:2002uc} 
by using various variants of the hybrid Monte Carlo (HMC) algorithm 
\cite{Duane:1987de}. The experience made with these algorithms was well summarized 
in a panel discussion at the Lattice 2001 Conference in Berlin \cite{Bernard:2002pd}. 
With Wilson-type fermions, major 
difficulties were encountered in trying to lower the fermion mass $m$ to values 
significantly smaller than half of the physical strange-quark mass $m_s$. In 
particular the cost of the simulations, at these masses and for those algorithms, 
increased by a power between 2 and 3 in $1/m$. Simulations with improved 
staggered fermions supplemented with the fourth-root trick were much faster, 
and much lighter quarks were already being simulated at that time~\cite{Bernard:2001av}. 
One of the main drawbacks of this formulation is that no 
local action with this determinant is known. 
It is thus not clear how to implement 
most of the usual quantum field theory machinery to properly define 
the continuum theory and obtain first-principle results. Since then,
an impressive numerical and theoretical amount of work has been done with 
this formulation. The last developments are reviewed at this conference by 
Sharpe~\cite{Sharpe:2006re}.

Over the last couple of years the situation  changed dramatically in this 
field, thanks to the development of the 
DD-HMC~\cite{Luscher:2003vf,Luscher:2003qa,Luscher:2004rx} and 
of the Hasenbusch-accelerated HMC with multiple-time scale 
integration~\cite{Hasenbusch:2001ne,Urbach:2005ji}. These 
algorithms allow for QCD simulations 
with light dynamical quarks which are much faster than before. 
This year, for the first time, large-scale simulations of two-flavour 
QCD with various actions, fine lattice spacings and quite large volumes 
have been performed with quark masses as light as $(m_s/5)$-$(m_s/4)$. Most of 
this talk is dedicated to summarizing the conceptual and numerical progress made thanks 
to these computations, with particular emphasis on the results obtained and the 
difficulties encountered at significantly smaller quark masses with respect to 
previous computations. I also review the first physical results
for pseudoscalar pion masses and decay constants and attempt to compare 
them with the corresponding expectations from chiral perturbation theory (ChPT). 

Full QCD simulation is a subject of very intense research in the lattice community.
Reviewing all contributions made over the past year or so in a single talk 
is not possible. The material chosen here reflects my personal taste and 
experience. I wish to apologize to those colleagues whose work it is not 
reviewed here.

\section{Cost of two-flavour simulations with Wilson-type fermions}\label{sec:cost}
Last year the authors of
Refs.~\cite{DelDebbio:2005qa,DelDebbio:2006cn,DelDebbio:2006qab}
extended the numerical experience with the DD-HMC 
algorithm~\cite{Luscher:2003vf,Luscher:2003qa,Luscher:2004rx},  
so to span across a wide range of parameter values. They 
simulated two-flavour QCD with quark masses as light as $m_s/4$, lattice 
spacings $a\sim 0.05$--$0.08$~fm and volumes with linear extensions of $L\sim 1.2$--$2.5$~fm.  
A crude cost formula that fits quite well their experience is~\cite{DelDebbio:2006cn}:
\be\label{eq:cost}
N_\mathrm{op} \sim k \left(\frac{\# \mathrm{confs}}{100}\right) 
                      \left(\frac{20\,\mbox{MeV}}{\overline m}\right)
		      \left(\frac{L}{3\, \mathrm{fm}}\right)^5
		      \left(\frac{0.1\; \mathrm{fm}}{a}\right)^6 \; 
                      \mathrm{Tflops}\times\mathrm{year}\; ,
\ee
where $a$ is the lattice spacing in fermi and $\overline m$ denotes the running 
sea-quark mass in the $\msbar$ scheme at the renormalization scale of $2$~GeV. 
For a volume of $2L\times L^3$ and for the Wilson gauge action, 
the pre-factor $k$ is $\sim0.03$ for Wilson fermions, while it is 
$\sim0.05$ for the Sheikholeslami--Wohlert (SW) action \cite{Sheikholeslami:1985ij}, 
with the coefficient $c_\mathrm{SW}$ fixed to the value determined non-perturbatively 
in Ref.~\cite{Jansen:1998mx}. When compared with analogous formulas presented at the Berlin 
2001 Lattice Conference, for instance the one proposed by Ukawa~\cite{Ukawa:2002pc}, 
the exponent of the quark mass is reduced from 3 to 1, that of the lattice 
spacing from 7 to 6, and the pre-factor $k$ is roughly 100 times smaller. A crucial ingredient in 
the DD-HMC algorithm, which allows for these performances, is the use of the Sexton--Weingarten 
multiple-time integration scheme~\cite{Sexton:1992nu}.

The Hasenbusch-accelerated HMC algorithm, with multiple-time
scale integration \cite{Hasenbusch:2001ne,Urbach:2005ji}, is being used 
in large-scale simulations of two-flavour QCD with SW~\cite{Gockeler:2006ns} 
and twisted-mass fermions~\cite{Jansen:2006rf}.
Even though the simulations carried out so far still span a moderate range of 
parameter values, the fact that these groups can already present first results 
at quark masses as light as  $(m_s/5)$--$(m_s/4)$ is very encouraging. In the last few 
months yet another algorithm, the rational HMC with multiple pseudofermion fields, 
has been proposed for dynamical simulations with light fermion 
masses~\cite{Clark:2006fx}. This algorithm and first tests carried out with Wilson fermions 
are reviewed at this conference by Clark~\cite{Clark:2006wq}.

The cost formula in Eq.~(\ref{eq:cost}) provides a simplified but clear summary 
of the progress made in full QCD simulations.
The reduced cost has allowed to simulate pion masses 
below the threshold of $m_\pi/m_\rho\sim0.5$ at fine lattice spacings and large volumes.
The most significant achievement reflected in this formula, i.e. the reduced exponent in the 
quark-mass dependence, tears down the ``Berlin Wall'' \cite{Bernard:2001av} 
and opens the way to simulations of lattices with pion masses as low as 
$250$--$300$ MeV. By inserting in Eq.~(\ref{eq:cost}) the values $\overline m=15 $~MeV, 
$a=0.05$~fm, $L=2.4$~fm, I obtain 
$N_\mathrm{op}\sim 0.8~\mathrm{Tflops}\times\mathrm{years}$ and 
$1.4~\mathrm{Tflops}\times\mathrm{years}$ for $100$ independent configurations
with Wilson and SW fermions, respectively. This means that a continuum extrapolation of simple 
observables could already be attempted at this mass with a machine of $2$--$4$ Tflops 
sustained!
 
\section{Spectral gap for the Dirac operator}\label{sec:bound}
For a given $\gamma_5$-Hermitian lattice Dirac operator $D_m$, 
it is convenient to consider the corresponding Hermitian operator 
\be
Q_m = \gamma_5 D_m\; , 
\ee
since it has the same determinant but its spectrum is real. For any gauge 
configuration on a finite lattice, the spectral gap and the spectral asymmetry 
of $Q_m$ are defined to be
\be
\begin{array}{lll}
\mu & = & \mbox{min}\,\Big\{|\lambda|\;\, \big|\; \lambda\; \mbox{is an eigenvalue of}\; Q_m\Big\}\\[0.25cm]
\eta & = & \frac{1}{2}\big\{N_+ - N_- \big\}\; ,
\end{array}
\ee
where $N_\pm$ are the numbers of positive and negative eigenvalues of $Q_m$.
The probability distributions of $\mu$ and $\eta$ are determined by the 
measure in the functional integral, and they are therefore properties of the 
regularized theory. Any decent simulation algorithm should reproduce 
them. Apart for its own physical interest in some cases (see below), the 
gap $\mu$ plays a crucial r\^ole for the stability of the simulations with 
HMC algorithms. In fact, if the probability for $\mu$ to be 
close to zero is not negligible, the HMC may run into molecular
dynamics integration instabilities, ergodicity problems, sampling 
inefficiencies, etc.~\cite{DelDebbio:2005qa}. In the large-volume regime and at 
large masses, physics considerations suggest that the average value 
of the spectral gap $\langle \mu \rangle$ is essentially proportional to the mass.
Naive thermodynamic arguments would also indicate that its squared width 
$\sigma^2=(\langle \mu^2 \rangle - \langle \mu \rangle^2)$ decreases 
proportionally with the inverse of the volume; see for example Ref.~\cite{DelDebbio:2005qa}.

The exact chiral symmetry preserved with Ginsparg--Wilson fermions 
ensures that  the gap is 
bounded from below when the mass $m$ is positive, i.e. $\mu\geq m$, and the asymmetry 
vanishes (see for example \cite{Niedermayer:1998bi}). Random matrix theory predicts 
the probability distributions of the low-lying eigenvalues of 
the Dirac operator for arbitrary values of $m$ \cite{Wilke:1997gf}. 
They reproduce the Leutwyler--Smilga sum
rules derived within the effective chiral theory. 
Some of their properties have been verified in quenched QCD with 
remarkable precision \cite{Giusti:2003gf}, and they are being verified in two-flavour 
QCD with increasing accuracy~\cite{Fukaya:2006xp}. 
By assuming the expressions derived in Ref.~\cite{Wilke:1997gf}, 
it is easy to show that for $u\equiv m\Sigma V\gg 1$ the first and second 
moments of the spectral gap distribution $p(\mu)$ are expected 
to go as 
\be\label{eq:rmt}
\frac{\displaystyle \langle \mu \rangle}{m} - 1 \rightarrow \frac{2}{u^2 \sqrt{\pi u}}\, , \qquad\qquad 
\frac{\displaystyle \langle \mu^2 \rangle - \langle \mu \rangle^2}{m^2} \rightarrow \frac{8}{u^4 \sqrt{\pi u}}\; .
\ee
These equations make it clear that chiral symmetry is ``freezing'' the 
fluctuations of the spectral gap, i.e. the width of its distribution 
decreases with a power of the 
volume much higher than that expected from naive thermodynamic arguments. 
When simulating smaller quark masses,  larger volumes 
can quickly stabilize dynamical simulations.
By inserting specific values in Eq.~(\ref{eq:rmt}), such 
as $\overline m=15 $~MeV, $L=2.4$~fm, $V=2L\times L^3$, and by assuming 
$\overline \Sigma = (250~\mathrm{MeV})^3$ in the $\msbar$ scheme at the 
renormalization scale of $2$~GeV, I obtain $u=10.33$ and 
\be\label{eq:musigC}
\frac{\langle \mu \rangle}{m} - 1 \sim 0.003\; , \qquad 
\frac{\sqrt{\langle \mu^2 \rangle - \langle \mu \rangle^2}}{m} \sim 0.011\; .
\ee
Also in twisted-mass QCD~\cite{Aoki:1983qi,Frezzotti:2000nk} the 
particular structure of the Dirac operator guarantees that the gap is 
bounded from below. In this case it may be interesting to study also the 
statistical distribution of the twist angle as a function of the quark mass.

\begin{figure}[t]
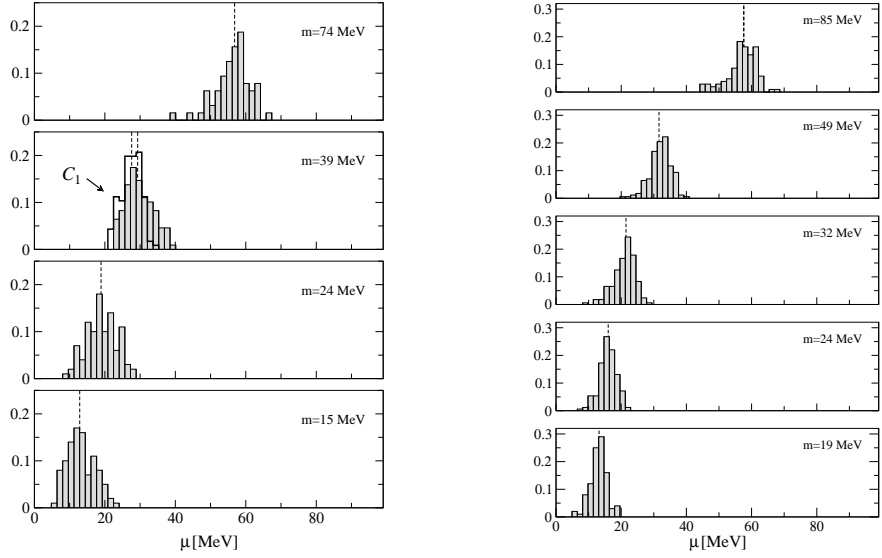

\vspace{0.5cm}

\begin{center}
\begin{tabular}{ccc}
\includegraphics[width=5.0cm]{32x24x24x24b5.60.minev.eps} & \qquad\qquad
\includegraphics[width=4.625cm]{48x24x24x24b5.30.minev.eps}
\end{tabular}
\caption{Normalized histograms of the spectral gap $\mu$. In the plot 
on the left they are obtained with the Wilson gauge and fermion action at 
$\beta=5.6$ on a $32\times24^3$ lattice, and for bare current quark masses $m$ 
as reported in the plots~\cite{DelDebbio:2005qa}. The darker line labeled 
by $C_1$ is obtained with the very same parameters, but with a volume of 
$64\times24^3$. The dotted vertical line is the median of the distribution. 
In the plot on the right there are analogous histograms for the Wilson gauge action and 
SW non-perturbatively improved fermions obtained at  
$\beta=5.3$ on a lattice $48\times24^3$ \cite{DelDebbio:2006qab}. 
}\label{fig:histmu}
\end{center}
\end{figure}

\begin{figure}[t]
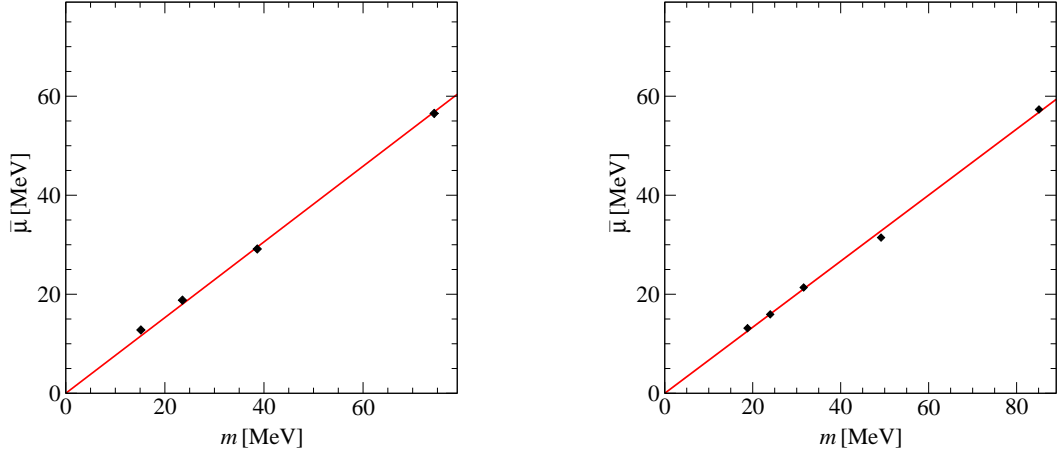

\vspace{0.5cm}

\begin{center}
\begin{tabular}{ccc}
\includegraphics[width=6.0cm]{32x24x24x24b5.60.avminev.eps} & \qquad\qquad
\includegraphics[width=6.0cm]{48x24x24x24b5.30.avminev.eps}
\end{tabular}
\caption{The median $\overline \mu$ of the spectral gap in MeV versus the bare current quark mass from the 
same numerical data as reported in Fig.~1.}\label{fig:avmu}
\end{center}
\end{figure}

The Wilson--Dirac operator (and related improved versions) breaks 
chiral symmetry explicitly. The above properties are thus
not expected to be valid. In particular the gap is not guaranteed 
to be bounded from below, and it is conceivable that $\mu$ could be 
much smaller than the current quark mass $m$  for some gauge-field 
configurations.  This year the progress in the algorithms allows for an empirical study 
of the spectral gap distribution at light-quark masses by numerical 
simulations~\cite{DelDebbio:2005qa}. The normalized histograms 
for $\mu$ obtained in Refs.~\cite{DelDebbio:2005qa,DelDebbio:2006qab} 
with the Wilson and with the SW-improved actions are shown in Fig.~\ref{fig:histmu}. 
At the volumes and masses considered, 
the distribution looks rather symmetric around the median $\overline \mu$, which 
is shifted toward smaller values when the quark mass decreases. 
The plots of the median versus the current quark mass shown 
in Fig.~\ref{fig:avmu} reveal that $\overline \mu$ is compatible 
with a linear function of the mass, the slope being roughly a number of 
order one at these lattice spacings. 

To date, the width of the distributions $\sigma$ has been 
studied at several volumes and masses with Wilson fermions only. 
Data indicate that $\sigma$ does not clearly depend on the 
quark mass and it scales as 
\be\label{eq:sigW}
\sigma \simeq \frac{a}{\sqrt{V}}\; ,
\ee
$V$ being the volume in physical units\footnote{Data at only one 
volume are available so far for SW fermions; see  Fig.~\ref{fig:histmu}. 
The width of the distribution shows a small trend to decrease 
with the mass \cite{DelDebbio:2006qab}.}. The proportionality with $a$ 
suggests that the width could result from short-distance fluctuations. 
An important conclusion that can be drawn from these results 
is that the spectrum of the Wilson operator at a given 
mass and for fine lattice spacings have, for practical purposes, a 
gap if the volume is large enough~\cite{DelDebbio:2005qa}. It would
be more than welcome to have an analytic control on the spectral gap 
distribution also in this case. A first step in this direction has already 
been taken this year in Ref.~\cite{Sharpe:2006ia}. 
By comparing the result in Eq.~(\ref{eq:sigW}) with the 
second formula in Eq.~(\ref{eq:rmt}), it can be noticed 
that the width of the distribution is much larger with 
Wilson fermions in the interesting ranges of parameter values. For example 
for $a\sim 0.072$, $L=2.4$~fm and for a lattice volume $2L\times L^3$, 
I get $\sigma\sim 1.7$~MeV, which is roughly one order of magnitude 
larger than the value reported in Eq.~(\ref{eq:musigC}).

The existence of a gap in the spectrum of the Wilson--Dirac operator is also 
one of the reasons why the new generation of algorithms can simulate 
these fermions so efficiently. The range of stability where
HMC algorithms can be safely applied can be defined, for instance, 
by requiring that $\overline \mu\ge 3\sigma$. By using the empirical 
fact that $\overline \mu\simeq Z m$ and $\sigma\simeq a/\sqrt{V}$, the bound can be written
as   
\be\label{eq:boundW}
m \ge \frac{3\sigma}{Z} \simeq \frac{3a}{Z\sqrt{V}}\; ,
\ee 
which clearly shows the dependence of the quark mass 
accessible to HMC simulations from the lattice spacing and size. 
Since it turns out (see below) that the ratio
$B = M_\pi^2/2m$ is practically independent of $m$, 
the previous bound can be written as 
\be
M_\pi\, L \ge \sqrt{3 \sqrt{2} a \frac{B}{Z}}\; .
\ee
By inserting the numerical values, the condition $M_P L\ge 3$
is sufficient for the bound to be satisfied at $a\le 0.09$~fm 
\cite{DelDebbio:2005qa,DelDebbio:2006qab}. A lattice with linear
extension $L\simeq2.4$~fm is more than sufficient for simulating a pion 
with $300$~MeV mass. Failing to fulfill the bound in Eq.~(\ref{eq:boundW})
could lead to instabilities in the particular HMC algorithm used, which 
in turn could even fake the presence of a phase transition in the theory.

\section{Two-flavour QCD results from the Schr\"odinger functional}
The ALPHA collaboration is continuing the long-term program of
non-perturbative renormalization of 
QCD~\cite{Luscher:1992an,Luscher:1991wu,Jansen:1995ck}.
The large-scale separation, which must be addressed, 
represents a formidable challenge for numerical simulations. 
The concept of an 
intermediate finite-volume renormalization scheme allows them to 
attack the problem on the lattice. The relation between
a hadronic quantity and a suitable observable defined in 
the finite-volume renormalization scheme is computed at 
low energy~\cite{Luscher:1992an}. 
The observable is then evolved non-perturbatively 
to higher scales using a recursive procedure \cite{Luscher:1991wu}. 
Eventually the perturbative regime is reached, 
and the matching with perturbation theory is straightforward 
(for a review see Ref.~\cite{Luscher:1998pe}).
The finite-volume renormalization scheme adopted is 
based on the Schr\"odinger functional (SF) ${\cal Z}$, i.e. the quantum propagation 
amplitude for going from some field configuration $C$ at time 
$x_0=0$ to another field configuration $C'$ at the time $x_0=T$:
\be
{\cal Z}[C',C] = e^{-\Gamma} = \int D[U,\psi,\bar\psi]\, e^{-S[U,\psi,\bar\psi]}\; ,
\ee
where $C$ and $C'$ depend on some parameters $\eta$ and $\nu$; see 
Ref.~\cite{Luscher:1992zx,Luscher:1993gh} for more details. 
The strong-coupling constant can, for instance, be defined 
as~\cite{Luscher:1993gh,DellaMorte:2004bc}
\be
{ \frac{\partial \Gamma}{\partial \eta}\Big|_{\eta=\nu=0}= \frac{k}{\bar g^2}}\, ,
\ee
where the normalization ${k}$ is chosen such that the tree-level value of 
$\bar g^2$ equals its bare value for all values of the lattice spacings.

\begin{figure}[t]
\setlength{\MiniPageLeft}{0.5\textwidth}
\setlength{\MiniPageRight}{\textwidth}\addtolength{\MiniPageRight}{-0.5\textwidth}
\begin{minipage}[t]{\MiniPageLeft}
\vspace{-1.25cm}

\includegraphics[width=7.5cm]{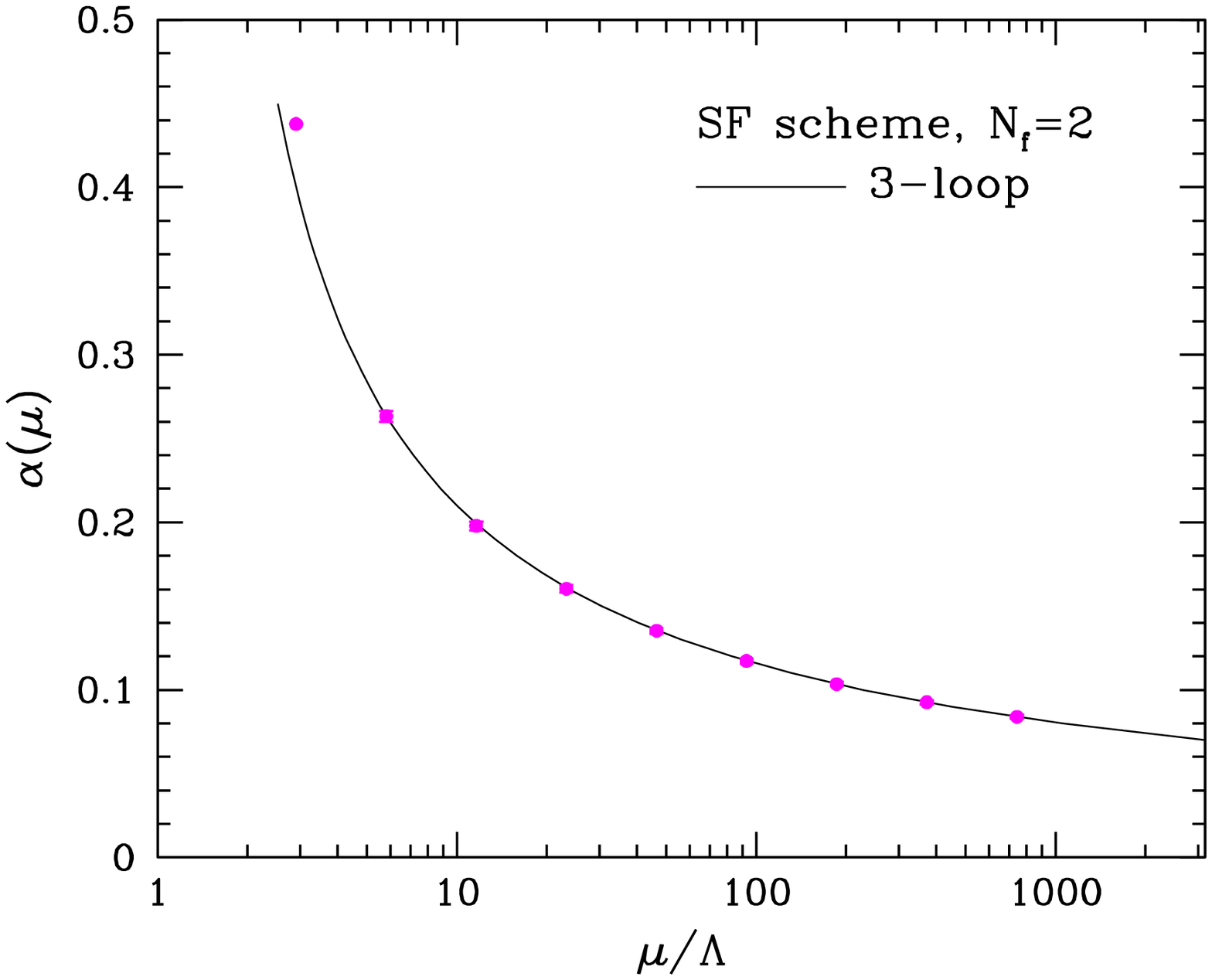} 
\end{minipage}
\begin{minipage}[t]{\MiniPageRight}
\vspace{-1.25cm}

\includegraphics[width=8.25cm]{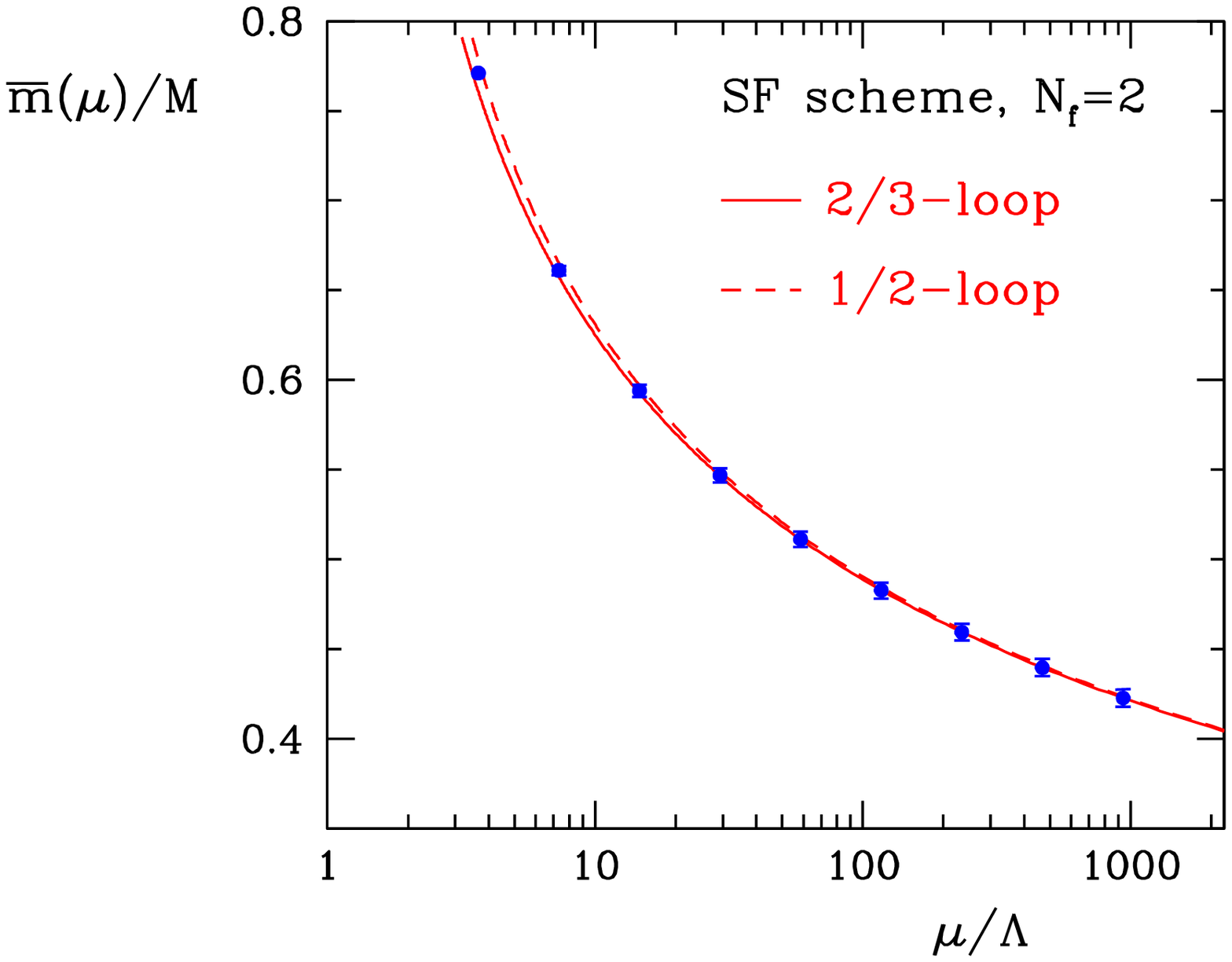} 
\end{minipage}
\vspace{-1.5cm}
\caption{Running of the strong coupling constant and of the quark mass in two-flavour QCD in 
         the Schr\"odinger functional scheme.}\label{fig:alpha}
\end{figure}

The ALPHA collaboration has recently completed the computation of the running 
of the coupling constant with two massless flavours~\cite{DellaMorte:2004bc}. 
They implemented the Schr\"odinger functional with a simple plaquette 
gauge action and with SW fermions, fixing the coefficient 
$c_\mathrm{SW}$ to the value determined non-perturbatively 
in Ref.~\cite{Jansen:1998mx}. The small discretization errors allowed them 
to safely extrapolate the results to the continuum limit with moderate errors,
and to obtain for the parameter $\Lambda$: 
\be
 -\log(\Lambda\, L_\mathrm{max}) = 1.09(7) \;\;\;\; \mathrm{at} \,\;\; u_\mathrm{max}=\bar g^2(L_\mathrm{max}) = 5.5\, . 
\ee
In the interval $u_\mathrm{max}=3.0$--$5.5$ their results can be 
parametrized as 
\be
-\log(\Lambda\, L_\mathrm{max}) = \frac{1}{2\, b_0\, u_\mathrm{max}} + 
\frac{b_1}{2\, b_0^2} \log(b_0\, u_\mathrm{max}) - 0.1612 + 0.0379\, u_\mathrm{max}\; .
\ee
The running of the coupling in the SF scheme as a function of 
$\mu/\Lambda$ is shown in the first plot of Fig.~\ref{fig:alpha}.
For $\alpha\le 0.2$ they observe an excellent agreement
with 3-loop perturbation theory, while at larger couplings the perturbative
approximation becomes inadequate quite rapidly. When compared with the running in 
the pure Yang--Mills theory~\cite{Luscher:1992an}, a clear $N_\mathrm{F}$ dependence is observed.
Thanks to their work, the energy dependence of the strong coupling in the SF scheme is now known over 
more than two orders of magnitude in two-flavour QCD.
The determination of $\Lambda$ in units of a physical hadronic scale
requires the computation of $L_\mathrm{max}$ in the same units. For lack of 
low-energy data, they used the Sommer scale~\cite{Sommer:1993ce} $r_0/a$ computed in 
Ref.~\cite{Gockeler:2004rp} and, by assigning to it the physical value of $r_0=0.5$~fm, 
they obtain $\Lambda^{(2)}_{\overline\mathrm{MS}}=245(16)(16)$~MeV. In full QCD,
$r_0$ tends to be a less convenient reference scale with respect to the quenched approximation: 
it requires large statistics at fine lattice spacing; it is not clear how to extrapolate 
its value from the simulated masses to the physical point; its physical value is not 
well determined since it cannot be measured directly in experiments. 
An important improvement in the determination
of $\Lambda$ can be achieved by computing $L_\mathrm{max}$ in units of the pion decay 
constant $F_\pi$ or the nucleon mass. Their most recent efforts in full QCD simulations 
with two flavours is motivated also by this goal~\cite{Meyer:2006ty}.

This year the ALPHA collaboration completed the computation 
of the factor that relates 
the running-quark mass in the Schr\"odinger functional scheme 
to the renormalization-group-invariant (RGI) one in two-flavour 
QCD~\cite{DellaMorte:2005kg}. In a mass-independent 
renormalization scheme, the relation between the RGI quark mass and 
the bare current mass is given by
\be
M = Z_M(g_0)\, m(g_0)\; .
\ee
The computation of $Z_M(g_0)$ can be split into two parts:
\be
Z_M(g_0) = \frac{M}{\overline m(\mu)}\,\frac{Z_A(g_0)}{Z_P(g_0,a\mu)}\; .
\ee
They computed the factor $M/\overline m(\mu)$, which is clearly 
regularization-independent 
but scheme-dependent, in the SF scheme over more than two orders of magnitude in energy, 
as shown in the right plot of Fig.~\ref{fig:alpha}. The matching factor 
$Z_A(g_0)/Z_P(g_0,a\mu)$ can be computed at low energy, and therefore 
no large energy differences are involved in the simulations. It 
is regularization- and scheme-dependent, and clearly depends also on the bare coupling. 
The ALPHA collaboration computed this factor in the 
SF scheme for the simple plaquette gauge action and for the SW non-perturbative 
improved fermions for a range of bare couplings \cite{DellaMorte:2005rd,DellaMorte:2005kg}. 
At this point the goal of computing the renormalized light-quark masses with controlled 
errors can be reached by matching these results with the computation of the bare 
current quark masses in large-volume simulations at light quark masses. 

\section{Effects of dynamical quarks in meson correlation functions}
One of the effects of quark--antiquark pair production is the coupling 
of multiparticle meson states to fermion bilinears. If the sea-quark 
masses are light enough, the first higher state in two-point 
correlation functions is expected to be a three-meson system, with energy 
roughly equal to $(M_0 +2\, M_\pi)$, where $M_0$ and $M_\pi$ are the masses
of the associated meson and of the pion made of sea quarks, respectively. 
It is clear that the lighter the sea-quark masses are, the smaller is the 
energy gap. As a consequence, the effect of the excited states should be more 
visible in an effective mass plot.

In Ref.~\cite{Allton:2004qq} the UKQCD collaboration studied the 
stability of the effective mass fit with regard to possible 
contaminations from higher states. In the range of quark masses simulated, 
the ground-state energy that they determine with a two-exponential fit is 
quite stable with regard to the fit details. For the higher-state energy
they find indicative values that turns out to be consistent with the expected three-meson 
state spectrum.
\begin{figure}[ht]
\begin{center}
\vspace{1.25cm}

\includegraphics[width=12.0cm]{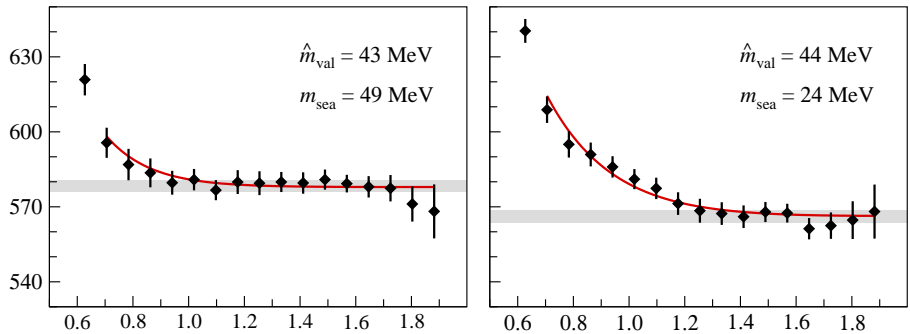}
\caption{Results for effective pion masses 
$M_\mathrm{eff}(t)$ in MeV as a function of
the time $t$ in units of fermi. They are computed with the SW non-perturbative 
improved action at $\beta=5.3$ on a volume of 
$48\times 24^3$~\cite{DelDebbio:2006cn}.}\label{fig:seaquarks}
\end{center}
\end{figure}

This year, much more precise results at much lighter quark 
masses are available~\cite{DelDebbio:2006cn}.  
Examples of effective-mass plots from the two-point function 
of non-singlet pseudoscalar densities are shown in 
Fig.~\ref{fig:seaquarks}. In the two plots the average valence-quark 
bare current masses $\hat{m}_{\rm val}$ and the meson masses $M_0$ (grey bands) are 
chosen to be nearly the same, while the bare sea-quark mass 
$m_{\rm sea}$ changes by a factor of $2$. The presence of higher-states 
contributions is clearly seen in the data, and a 
statistically significant quark-mass dependence is observed~\cite{DelDebbio:2006cn}. 
The effect of higher states 
becomes more pronounced when the sea-quark mass become lighter. The solid line
is a fit of the form
\be\label{eq:twos}
M_\mathrm{eff}(t) = M_0 + c e^{-2\,M_{\pi} t}+ \cdots\; ,
\ee
where $M_0$ and $c$ are free parameters, and $M_{\pi}$ is extracted from the 
pseudoscalar correlation function made of sea-quarks only. The 
$\chi^2/\mathrm{d.o.f.}$ of the fits indicates that the two-state formula 
in Eq.~(\ref{eq:twos}) is compatible with the data in the time range considered. 
In Ref.~\cite{DelDebbio:2006cn} analogous effects are also 
observed in the vector channels.

Apart for its own physical interest, it is quite clear 
that at light-quark masses the presence of multimeson states
will complicate the extraction of ground-state energies from the simulated 
data. The computation of hadron masses may require accurate data at larger time 
separations than was the case in quenched QCD.
\begin{figure}[t]
\setlength{\MiniPageLeft}{0.6\textwidth}
\setlength{\MiniPageRight}{\textwidth}\addtolength{\MiniPageRight}{-0.6\textwidth}
\begin{minipage}[t]{\MiniPageLeft}
\vspace{-0.5cm}

\includegraphics[width=7.0cm]{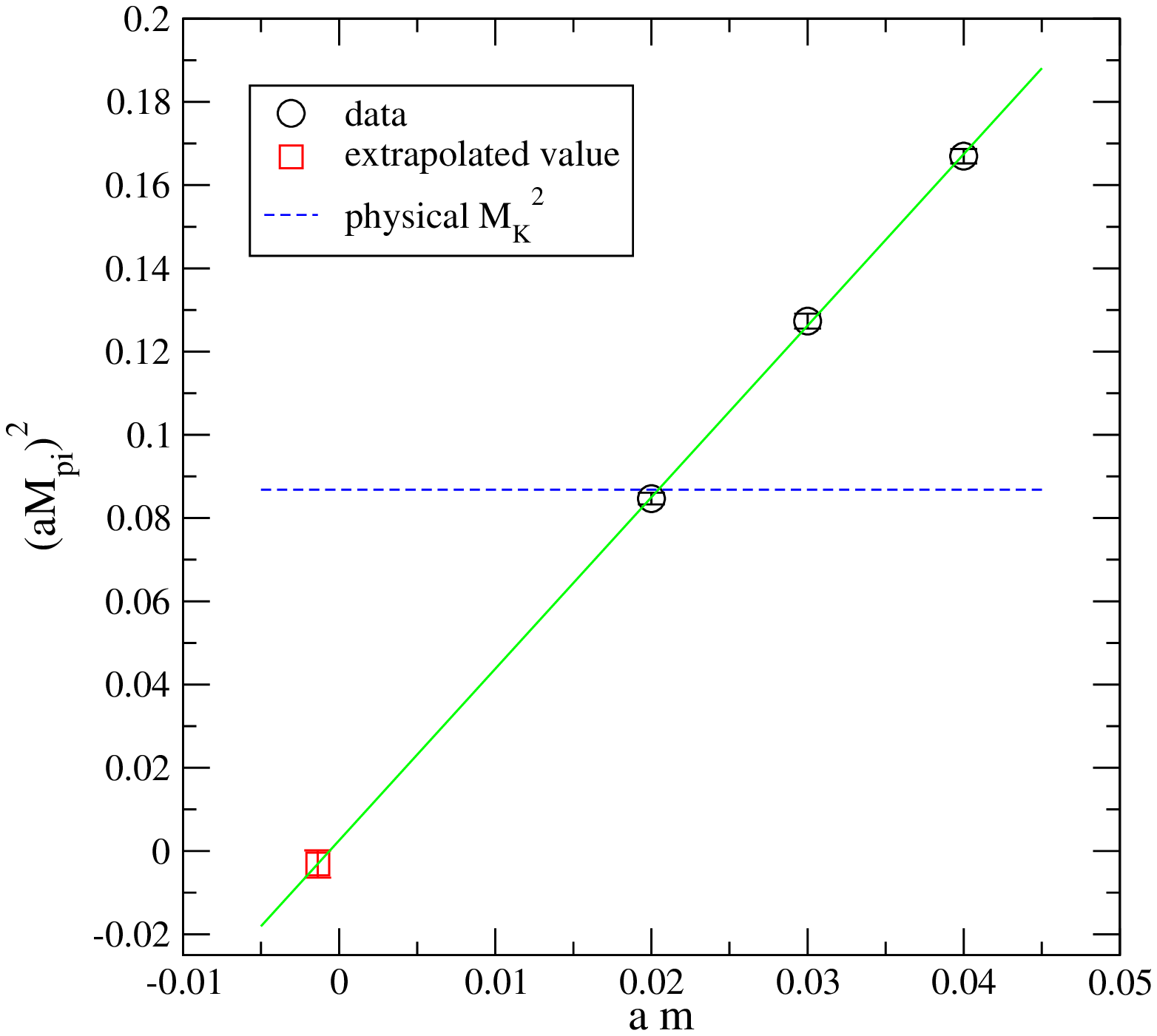}
\end{minipage}
\begin{minipage}[t]{\MiniPageRight}
\vspace{0.5cm}

\includegraphics[width=6.0cm]{fp_lin_0411006.eps}
\end{minipage}
\caption{Quark-mass dependence of the pion mass square and decay constant computed
         with domain-wall fermions on a volume $32 \times 16^3$, with a lattice 
         spacing of $a \sim 0.12$~fm~\cite{Aoki:2004ht}. }\label{fig:BNL}
\end{figure}

\section{Pseudoscalar meson masses and decay constants from 
         two-flavour simulations}\label{sec:rawdata}
In the recent past, pion masses and decay constants were
computed in two-flavour QCD at quark masses $m\ge m_s/2$ by 
several collaborations~\cite{Eicker:1998sy,AliKhan:2001tx,
Allton:2001sk,Aoki:2002uc,Allton:2004qq,Aoki:2004ht}. Data 
from Ref.~\cite{Aoki:2004ht} generated with domain-wall fermions  
on lattices of size $32 \times 16^3$ with a lattice 
spacing of $a \sim 0.12$~fm are shown 
in Fig.~\ref{fig:BNL}. Be it for the pion mass square 
or for the decay constant, they find a remarkable 
linear behaviour in the range $(m_s/2)$--$m_s$. 
Similar results have been obtained by the other collaborations 
with different gluon and fermions actions. After this experience with 
two-flavour QCD, the RBC and the UKQCD collaborations decided 
to move on and simulate $2+1$ flavours with domain-wall fermions. 
The first preliminary results have already been presented at this 
conference. The reader can find details of these simulations in their 
talks~\cite{Mawhinney:2006lat,Maynard:2006lat,Lin:2006lat,Allton:2006ax}. 
The PACS-CS 
collaboration is working hard to implement a combination of 
DD-HMC  and PHMC algorithms to simulate QCD with $2+1$ flavours
with the SW non-perturbative improved fermions and with light up and 
down quark masses. The first experience of this remarkable effort 
have already been reported at this conference in 
Refs.~\cite{Ukawa:2006pre,Ishikawa:2006pb,Kuramashi:2006np}
\begin{figure}[t]
\vspace{0.625cm}

\begin{center}
\includegraphics[width=12.0cm]{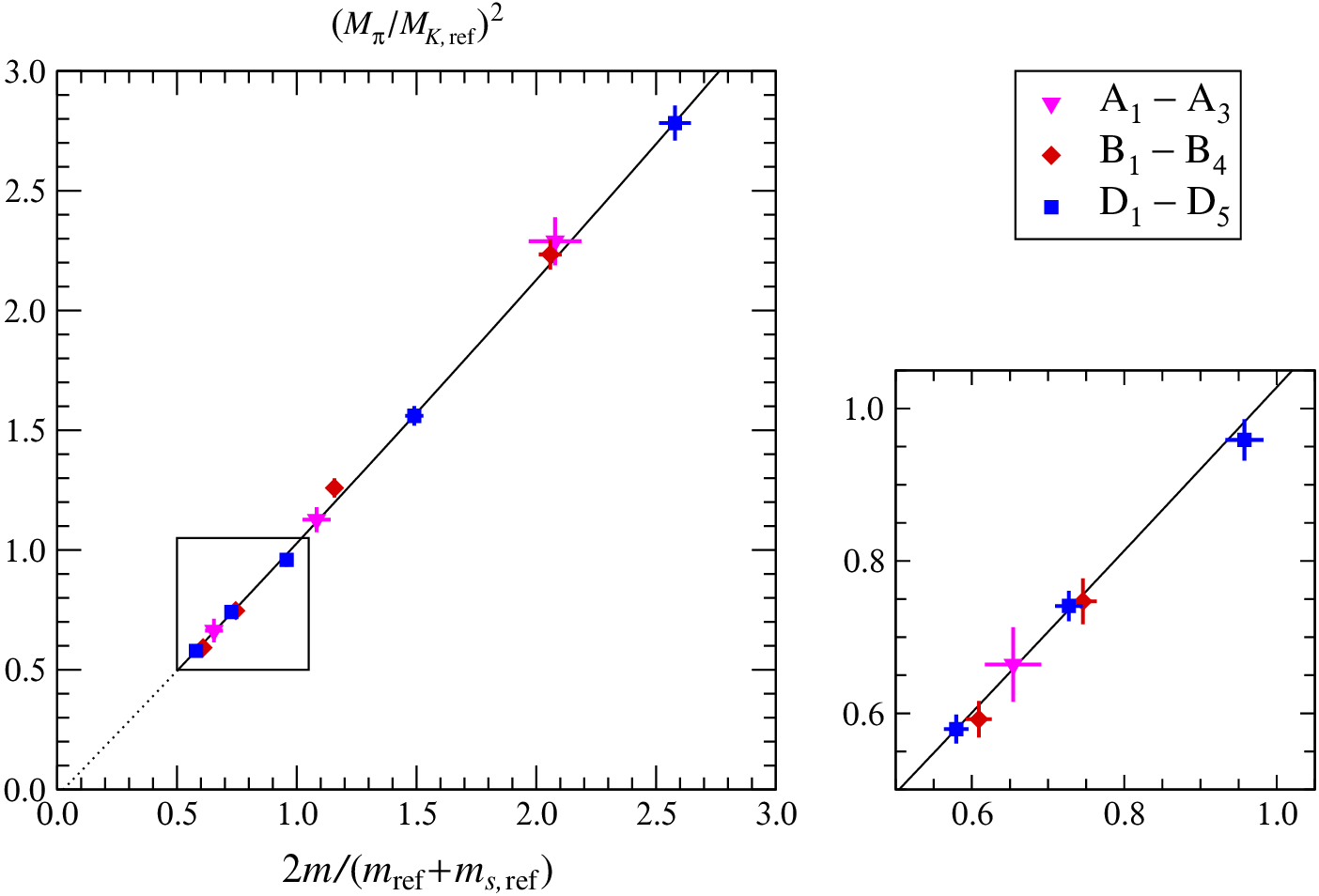}
\end{center}
\vspace{0.5cm}

\setlength{\MiniPageLeft}{0.5\textwidth}
\setlength{\MiniPageRight}{\textwidth}\addtolength{\MiniPageRight}{-0.5\textwidth}
\begin{minipage}[t]{\MiniPageLeft}
\setlength{\tabcolsep}{.25pc}
\begin{tabular}{cccrr}
    & Lat & $k$ &$N_\mathrm{trj}$ & $N_\mathrm{conf}$\\
\hline
 {          W/W}            & $A_1$ & 0.15750 & 6400 & 64 \\
 {       $V\,a^{-4}=32\times 24^3$} & $A_2$ & 0.15800 & 10900& 109\\
 {       $\beta = 5.6$}     & $A_3$ & 0.15825 & 10000& 100\\
\hline
 {       W/W}               & $B_1$ & 0.15410 & 5000 & 100\\
 {       $V\,a^{-4}=64\times 32^3$} & $B_2$ & 0.15440 & 5050 & 101\\
 {       $\beta=5.8$}       & $B_3$ & 0.15455 & 5200 & 104\\
                            & $B_4$ & 0.15462 & 5100 & 102\\
\hline
\end{tabular}
\end{minipage}
\begin{minipage}[t]{\MiniPageRight}
\vspace{-1.75cm}

\setlength{\tabcolsep}{.25pc}
\begin{tabular}{cccrr}
    & Lat & $k$ &$N_\mathrm{trj}$ & $N_\mathrm{conf}$\\
\hline
 {      W/SW}              & $D_1$ & 0.13550 & 5200 & 104\\
 {      $V\,a^{-4}=48\times 24^3$} & $D_2$ & 0.13590 & 5130 & 171\\
 {      $\beta=5.3$}       & $D_3$ & 0.13610 & 5040 & 168\\
                            & $D_4$ & 0.13620& 5040 & 168\\
                            & $D_5$ & 0.13625& 5040 & 169\\
\hline
\end{tabular}
\end{minipage}
\caption{Parameters of the lattices generated in Refs.~\cite{DelDebbio:2006cn,DelDebbio:2006qab} 
with the Wilson gauge action and Wilson ($W/W$) and SW non-perturbative improved ($W/SW$) fermions
are given in the tables. $k$ is the hopping parameter, 
$N_\mathrm{trj}$ is the number of HMC trajectories generated after thermalization,
and $N_\mathrm{conf}$ is the number of independent configurations selected to compute the observables. 
The dependence of the square of the pion mass $M_\pi$ 
on the sea-quark current mass $m$ in units of the same quantities at the reference point is shown in the plots.
The solid curve is a quadratic least-squares fit (with constant term) of all data points. The plot on 
the right is a blow up of the region enclosed by the little box.}\label{fig:mpius}
\end{figure}

This year for the first time we have a quite large amount of results 
obtained in two-flavour QCD at quark masses as low as $(m_s/5)$--$(m_s/4)$. 
The richest set of data has been accumulated with the DD-HMC algorithm 
in Refs.~\cite{DelDebbio:2006cn,DelDebbio:2006qab}. Tables with the 
actions implemented, lattice parameters, and number of 
configurations generated are shown in Fig.~\ref{fig:mpius}.
The authors supplement the two-flavour theory with a 
quenched strange quark. They define the reference 
point to be where the mass of the $\pi$, $K$ and $K^*$ mesons satisfy 
\be
\frac{M_{K,\mathrm{ref}}}{M_{K^*,\mathrm{ref}}} = 0.554\, \qquad \qquad 
\frac{M_{\pi,\mathrm{ref}}}{M_{K,\mathrm{ref}}}=0.85\; .
\ee
The lattice spacing $a$ is then fixed by setting the 
value $M_{K,\mathrm{ref}}=495$~MeV~\cite{DelDebbio:2006cn}. 
This procedure extends similar ideas already implemented in the quenched 
approximation~\cite{Allton:1996yv,Heitger:2000ay}. For the three sets of 
lattices  $A$, $B$ and $C$ (see table in Fig.~\ref{fig:mpius}) they obtain
$a=0.0717(15)$, $0.0521(7)$ and $0.0784(10)$, respectively. These values
are significantly smaller that those reported 
in Ref.~\cite{Luscher:2005mv}, determined by fixing $a$ from the Sommer scale. 
As a consequence the pion and the quark masses in physical units quoted here are significantly 
larger than in Ref.~\cite{Luscher:2005mv}. 
\begin{figure}[t]
\setlength{\MiniPageLeft}{0.45\textwidth}
\setlength{\MiniPageRight}{\textwidth}\addtolength{\MiniPageRight}{-0.45\textwidth}
\begin{minipage}[t]{\MiniPageLeft}
\vspace{-6.0cm}

\setlength{\tabcolsep}{.25pc}
\hspace{0.25cm}\begin{tabular}{ccccc}
    & $a \mu$ & $N_\mathrm{trj}$ \\
\hline
 {       tlSym/tm}                     & 0.0150 & 5000 \\
 {       $V\,a^{-4}=48\times 24^3$}            & 0.0100 & 5000 \\
  {       $\beta = 3.90$}              & 0.0064 & 5000 \\
                                       & 0.0040 & 5000 \\
\hline
 {       tlSym/tm}                     & 0.0060 & 500  \\
 {       $V\,a^{-4}=64\times 32^3$}            & 0.0030 & 2200 \\
  {      $\beta = 4.05$}              &        &      \\
\hline
\end{tabular}
\end{minipage}
\begin{minipage}[t]{\MiniPageRight}

\includegraphics[width=6.5cm]{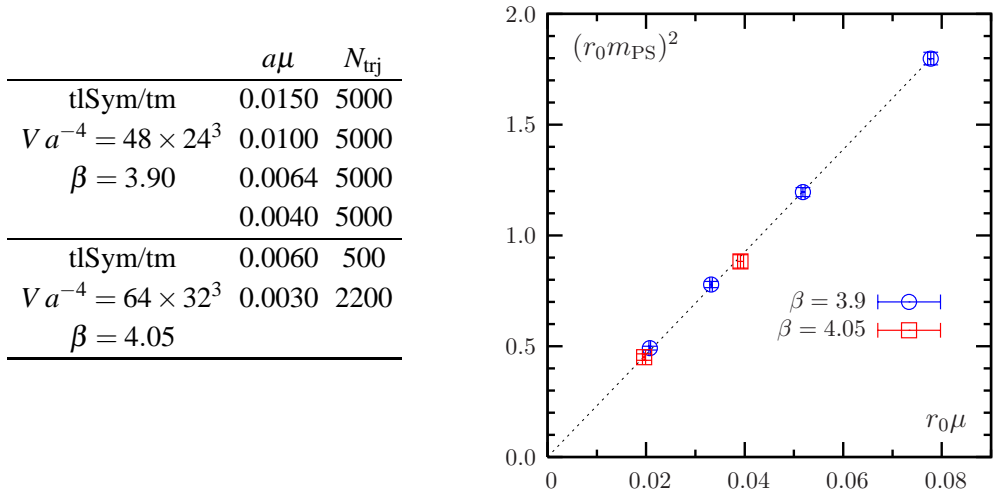}

\end{minipage}
  \caption{Current status of the simulations with the tree-level Symanzik improved 
gauge action (tlSym) and twisted-mass fermions at maximal twist reported 
in Ref.~\cite{Jansen:2006rf}. The bare quark mass $a \mu$ and the number 
of trajectories generated after thermalization are listed in the table. 
The dependence of the square of the pion mass $M_\pi$ on the sea-quark 
twisted mass $\mu$ is shown in the plot. Masses are expressed in 
units of $r_0$ computed at the lightest simulation point.}
  \label{fig:tmQCD}
\end{figure}

Once the lattice is calibrated, results from the various sets of 
simulations can be compared. In Fig.~\ref{fig:mpius} the ratio 
$(M_\pi/M_{K,\mathrm{ref}})^2$ is shown as a function of the 
corresponding ratio of current quark masses. The plots reveal several 
remarkable properties of the results. Data generated with two 
different discretizations and three lattice spacings lie on the 
same ``universal'' curve within the statistical fluctuations. This 
supports the fact that the discretization effects in the relation between 
the pseudoscalar-meson-mass squared and the current quark mass are small. 
This fact could be explained with the observation that $O(a)$ effects 
are absent at leading order in chiral perturbation theory if the current quark mass
is used~\cite{Sharpe:1998xm}. The second remarkable property is the 
linear behaviour of the data over such a wide range of quark masses.
There is a visible curvature towards the larger masses, but the 
coefficient of the quadratic term in the empirical fit (solid line) 
is small. Moreover in the range $(M_\pi/M_{K,\mathrm{ref}})\le 1.1$, the results 
are well represented by a straight line through the origin. 

This year the QCDSF-UKQCD collaboration supplemented their  set of data 
generated with the SW non-perturbative improved action
at $\beta=5.29$ with two new points with quark masses well below $m_s/2$~\cite{Gockeler:2006ns,Gockeler:2006vi}. 
Even though they are based on a still limited statistics, their findings for $M_\pi^2$ 
are compatible with the previous observations. The ETM collaboration is simulating two-flavour QCD 
with the tree-level Symanzik improved gauge action (tlSym) and twisted-mass fermions at maximal twist. 
At this conference they presented 
data at fine lattices and small quark masses~\cite{Jansen:2006rf}. 
A summary of the parameters of their simulations and of their results for the 
pseudoscalar mass square are reported and shown in Fig.~\ref{fig:tmQCD}.
The value of $r_0$ used to plot the data is the one obtained at their lightest
simulation point. Data collected so far are well compatible with a 
linear behaviour of $M_\pi^2$ versus the quark mass.

\begin{figure}[t]
\vspace{1.25cm}

\begin{center}
\includegraphics[width=12.0cm]{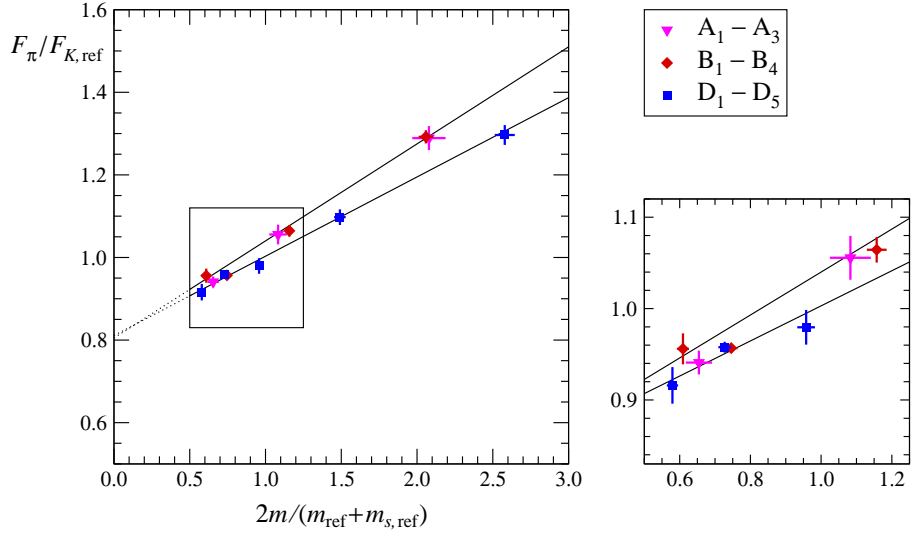}
\end{center}
\caption{Dependence of the pion decay constant $F_\pi$ on the sea-quark mass $m$
in units of the same quantities at the reference point as 
obtained in Ref.~\cite{DelDebbio:2006cn,DelDebbio:2006qab}. The solid curves 
are linear fits of the data points from the $A$ and $B$ lattices (upper line) and of 
the points from the $D$ lattices (lower line).The plot on 
the right is a blow up of the region enclosed by the little box.}\label{fig:fpius}
\end{figure}

Data generated in Refs.~\cite{DelDebbio:2006cn,DelDebbio:2006qab} for pion decay 
constant $F_\pi$ are shown in Fig.~\ref{fig:fpius}. All data sets are statistically 
compatible with a linear behaviour in the range of mass explored. The results of the 
$D$ lattices turn out to be quite different from those of the $A$ and $B$ lattices.  
Although the two lines are visibly different, the fitted values of their slopes, 
$0.235(11)$ and $0.192(11)$, deviate from each other by less than $3$ times the combined 
statistical error. The statistical significance of the effect is thus not conclusive. 
A similar linear behaviour in $F_\pi$ has been observed also in the data of 
the ETM collaboration~\cite{Jansen:2006rf}.  

\section{Finite-volume corrections for pseudoscalar meson masses}\label{sec:finvol}
Finite-volume effects are expected to be negligible in hadron masses 
if the linear extension $L$ of the lattice is much 
larger than the cloud of virtual particles surrounding them. 
In this regime of volumes, i.e.  $F L \gg 1$ and $M_\pi L \gg 1$, 
the leading finite-volume corrections to pion masses can be estimated 
in ChPT. At the next-to-leading order
(NLO)~\cite{Gasser:1986vb,Gasser:1987ah,Gasser:1987zq} 
\be\label{eq:NLOFV}
M_\pi^2(L) =  M_\pi^2\left\{ 1 + \frac{1}{2}\frac{M^2}{(4\pi F)^2}\, g_1(M) \right\}
\ee
where
\be
g_1(M) = 4 \sum_{\{n_k\}}\,\!\!\! ' \frac{1}{M |q_n|} K_1(|q_n| M)\; .
\ee
$M^2=2\, B \overline m$ is the pion mass at leading order, 
$B$ and $F$ are the leading-order low-energy constants, 
$ \sum_{\{n_k\}}\,\!\!\!\!\!\!\!\!\!\!\!\!\!\!\! '\;\;\;\;$ denotes the sum over all 
three-dimensional vectors of integers with the exception 
of $n=(0,0,0)$, $q_n^2 = \sum_{k=1}^3 (n_k L)^2$, and $ K_1$  is a modified Bessel 
function. These corrections are 
exponentially small in $M_\pi L$, i.e. the leading exponential in 
Eq.~(\ref{eq:NLOFV}) is given by
\be\label{eq:NLOFVII}
\displaystyle \frac{M_\pi(L) - M_\pi}{M_\pi} =  \displaystyle \frac{M^2}{(4\pi F)^2} 
                                     { \frac{3 \sqrt{2\pi}}{(M L)^{3/2}} e^{-M L}}\; .
\ee
Once the (infinite) volume values of $F_\pi$ and $M_\pi$ are known, 
Eq.~(\ref{eq:NLOFV}) gives a parameter-free and model-independent estimate of the 
leading corrections at asymptotically large volumes. An analogous formula can be written 
for the decay constant. Whether this regime has been reached 
for the lattices currently simulated in two-flavour QCD still needs  
to be confirmed.  Other effects such as the squeezing of the meson
wave function~\cite{Fukugita:1992jj} cannot be taken into account in ChPT, 
where the pion is a ``point-like'' particle, and therefore we do not have 
any analytic handle on them. 

\begin{figure}[t]
\setlength{\MiniPageLeft}{0.6\textwidth}
\setlength{\MiniPageRight}{\textwidth}\addtolength{\MiniPageRight}{-0.6\textwidth}
\begin{minipage}[t]{\MiniPageLeft}
\vspace{0.25cm}

\setlength{\tabcolsep}{.2pc}
\begin{tabular}{cccccc}
  $\beta$  &    $K$   & $T\times L^3$    & $a M_\pi(L)$   & $M_\pi L$& Ref.\\
\hline
     5.6   &  0.1575  & $32 \times 12^3$& 0.3576(89) & 3.31(2)& \cite{Orth:2005kq}\\
           &          & $32 \times 14^3$& 0.3048(44) & 3.86(3)& \cite{Orth:2005kq}\\
           &          & $32 \times 16^3$& 0.2806(35) & 4.41(3)& \cite{Orth:2005kq}\\
           &          & $40 \times 24^3$& 0.2765(26) & 6.61(4)& \cite{Orth:2005kq}\\
           &          & $32 \times 24^3$& 0.2744(21) & 6.61(4)& \cite{DelDebbio:2006qab}\\
\hline
     5.6   &  0.1580  & $32 \times 16^3$& 0.233(5)   & 3.15(3)& \cite{Orth:2005kq}\\
           &          & $32 \times 16^3$& 0.242(4)   & 3.15(3)& \cite{Luscher:2004rx}\\
           &          & $40 \times 24^3$& 0.1991(33) & 4.73(4)& \cite{Orth:2005kq}\\
           &          & $32 \times 24^3$& 0.1969(16) & 4.73(4)& \cite{DelDebbio:2006qab}\\
\hline
\end{tabular}
\end{minipage}
\begin{minipage}[t]{\MiniPageRight}
\vspace{0.25cm}

\includegraphics[width=6.0cm]{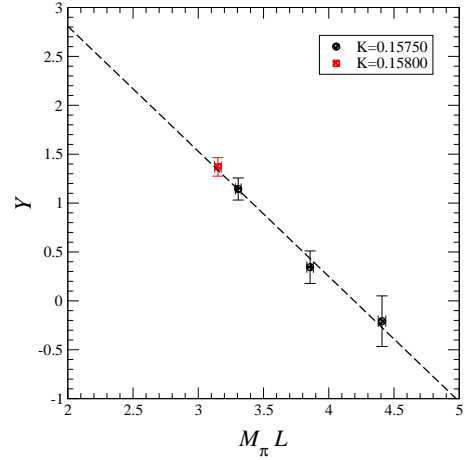}
\end{minipage}
\caption{Meson masses computed with the simple plaquette gauge action and Wilson fermions 
at $\beta=5.6$ for various lattice sizes. The value of $(a\, M_\pi)$ is 
obtained by combining data at the largest spatial volume.
In the plot the quantity $Y$, defined in Eq.~(\protect\ref{eq:Y}), is shown as a 
function of $M_\pi L$. The dashed line is a linear fit to the data points.}\label{eq:sesam}
\end{figure}

Thanks to their universality, finite-size effects can be estimated by simulating 
lattices with different volumes at a single (fine) lattice spacing. 
Discretization effects add small corrections, 
which can be neglected to a first approximation. A careful study for the  
pion mass has been carried out in Ref.~\cite{Orth:2005kq}. This 
year their data can be supplemented by the results from 
Refs.~\cite{Luscher:2004rx,DelDebbio:2006qab}, generated 
with the same simple-plaquette gluon action and with Wilson fermions 
at the very same mass. The values of pion masses at volumes 
that satisfy the stability bound in Section~\ref{sec:bound} are reported in the 
table of Fig.~\ref{eq:sesam}. 
To compare the ChPT prediction in Eq.~(\ref{eq:NLOFV}) with numerical data,
I make the working assumption that finite-volume corrections 
at the largest volumes are negligible
within the statistical errors. This assumption is compatible with 
Eq.~(\ref{eq:NLOFV}). In Fig.~\ref{eq:sesam} the 
quantity\footnote{Comparable data from
the two collaborations have been combined. It is reassuring that for two volumes and 
at two different masses, the two groups obtain compatible results, within the statistical 
errors, by generating the gauge configurations with two different algorithms.}
\be\label{eq:Y}
Y= \log\left[\displaystyle \frac{M_\pi(L)-M_\pi}{M_\pi}\times 
                           \frac{(M_\pi L)^{3/2}}{3\sqrt{2\pi} (a M_\pi)^2}\right]  
\ee
is shown in logarithmic scale as a function of $M_\pi L$. The linearity of the data agrees well with an 
exponential behaviour of the form $e^{-M_\pi L}$. The NLO ChPT correction
in Eq.~(\ref{eq:NLOFV}), however, underestimates the pre-factor by roughly one order of 
magnitude at these volumes and quark masses. This means that 
a volume with a linear extension of $L\sim 1.2$~fm is far too small for ChPT 
to apply. Even though we do not have a quantitative explanation for it, the mismatch 
between the data and ChPT is not unexpected \cite{Fukugita:1992jj}.
In such a small volume the pion wave-function is most likely modified 
significantly, and treating the pion as a point-like particle 
is an approximation too crude. What is maybe more 
important in this regard is to find the lattice sizes where finite-volume 
effects at the masses currently simulated match the asymptotic correction 
expected in the chiral effective theory, or where these corrections are negligible 
with respect to the statistical errors. This is the minimal requirement for the 
data to be useful, and eventually for being compared with ChPT.
It is most likely that this goal can be reached in the very near future. Similar 
considerations apply to $F_\pi$.

\section{Two-flavour QCD at fixed topology}
This year the JLQCD collaboration started an ambitious project 
of simulating two-flavour QCD with Neuberger's fermions at fixed 
topology~\cite{Hashimoto:2006rb,Matsufuru:2006xr,Kaneko:2006pa,Fukaya:2006xp}.
The global topological charge is frozen by supplementing the Iwasaki 
gluon action with the extra Boltzmann weight in the functional integral
\be\label{eq:JLQCDadd}
\frac{\mathrm{det}\; H^2_\mathrm{W}}{\mathrm{det}\, (H^2_\mathrm{W} + \mu^2)}\; ,
\ee
where $H_\mathrm{W}$ is the Hermitian--Wilson operator with a large 
negative mass, and $\mu$ is a real parameter~\cite{Fukaya:2006vs}. 
This is equivalent to introducing 
two additional flavours of Wilson fermions with unphysical large 
negative mass, and two additional twisted-mass ghosts with large mass 
$\mu$. Since these extra fields have masses of the order of the cut-off, 
they modify the theory in the ultraviolet only, and thus become 
irrelevant in the continuum limit. For the light physical quarks, JLQCD 
chooses to work with Neuberger fermions~\cite{Neuberger:1997fp}. Earlier 
studies in the quenched approximation showed that 
this modified gluon action prevents the near-zero modes of 
$H_\mathrm{W}$ from reaching zero~\cite{Fukaya:2006vs}. 
It thus fixes the global topological 
charge as defined via the index of the corresponding Neuberger operator.
In this way data at fixed topology can be generated efficiently: the topological charge 
does not have to be computed for each configuration, and only those 
with the desired topology are produced. An important technical advantage 
is that the discontinuities that appear in the HMC Hamiltonian 
when a near-zero mode of $H_\mathrm{W}$ passes through zero are avoided.
Simulations with a plain HMC algorithm are thus feasible.
The cluster decomposition property of the underlying local quantum 
field theory suggests that fixing the global topological charge 
should have harmless effects on physical observables at asymptotically 
large volumes. However, unlike what happens in the full theory, 
the suppression of finite-size effects should be power-like in $L$ 
rather than exponential. At the volumes accessible so far in numerical 
simulations, these effects can be relevant and a careful analysis 
of them is needed.
\begin{figure}[t]
\setlength{\MiniPageLeft}{0.5\textwidth}
\setlength{\MiniPageRight}{\textwidth}\addtolength{\MiniPageRight}{-0.5\textwidth}
\begin{minipage}[t]{\MiniPageLeft}
\vspace{0.25cm}

\setlength{\tabcolsep}{.2pc}
\begin{tabular}{ccccc}
    & $a m$ & $N_\mathrm{trj}$ \\
\hline
                                      & 0.100 & 3590 \\
  $IW^{*}/N$                          & 0.070 & 3500 \\
  $\beta = 2.30$                      & 0.050 & 3500 \\
  $V a^{-4} =32\times 16^3$           & 0.035 & 4150 \\
     $\mu=0.2$                        & 0.025 & 4320 \\
   $\nu=0$                            & 0.015 & 2150 \\
\hline
\end{tabular}

\end{minipage}
\begin{minipage}[t]{\MiniPageRight}
\vspace{0.25cm}

\hspace{-0.5cm}\includegraphics[width=8.0cm]{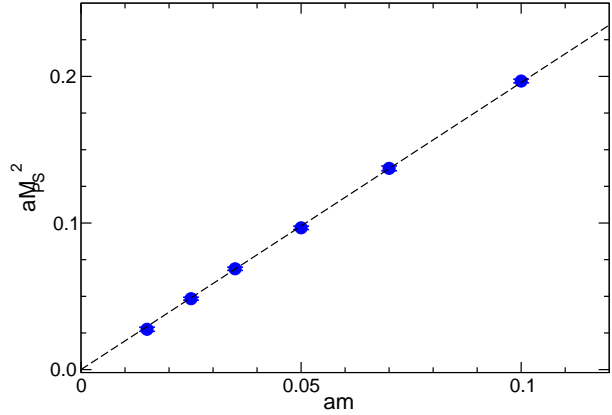}
\end{minipage}
\caption{The table shows the lattice parameters of the main JLQCD simulations
with the Iwasaki gluon action, with the extra weight defined 
in Eq.~(\protect\ref{eq:JLQCDadd}), and Neuberger's fermions ($IW^{*}/N$):
$am$ is the bare quark mass and $N_\mathrm{trj}$ is the number of trajectories
generated. The plot represents the quark-mass dependence of the pion mass square; 
the dashed line is a linear fit of the data.}\label{fig:JLQCD}
\end{figure}

The JLQCD collaboration generates the gauge configurations with the 
Hasenbusch-accelerated HMC algorithm with multiple-time scale 
integration \cite{Matsufuru:2006xr}. The lattice volume is
$32\times 16^3$, $\beta=2.30$ and so far the topological charge is $\nu=0$. 
With this set-up they expect to have a lattice spacing of roughly $0.12$~fm 
and a linear size extent of $L\sim 1.9$~fm. The list of bare quark masses 
considered is given in the table of Fig.~\ref{fig:JLQCD}. They should be in 
the range $(m_s/6)$--$m_s$. The first preliminary results that JLQCD has obtained 
for the pion mass squared are shown in the plot of Fig.~\ref{fig:JLQCD} as a function of 
the bare quark mass. Also in their data a remarkable linear behaviour is observed with  
no statistically significant deviation from a simple linear fit (dashed line). 
Before drawing any physics conclusions from these preliminary results, 
more simulations at several volumes and topologies are needed for 
a careful study of finite-size effects~\cite{Kaneko:2006pa}.

\begin{figure}[t]
\begin{center}
\includegraphics[width=8.0cm]{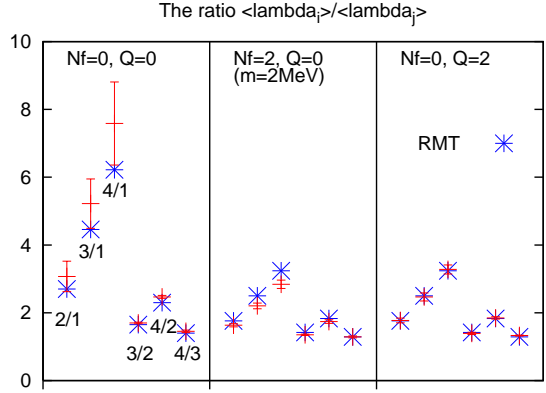}
\caption{Ratios of low-lying eigenvalues  $\lambda_i/\lambda_j$ 
(denoted by ``$i/j$'') of the Neuberger operator for 
$(N_\mathrm{F}, Q)=(0,0)$, $(2,0)$, $(0,2)$ (crosses). The random-matrix-theory 
predictions are also indicated (stars).}\label{fig:fukaya}
\end{center}
\end{figure}

JLQCD performed also a test run in the $\epsilon$-regime of QCD, 
on a lattice of $32\times 16^3$, $\beta=2.35$ and $am=0.002$.
The statistics accumulated is still limited to $1400$ trajectories,
but preliminary results for ratios of eigenvalues and pseudoscalar correlation
functions were already presented in Ref.~\cite{Fukaya:2006xp}. A 
comparison with random-matrix-theory predictions is shown in 
Fig.~\ref{fig:fukaya}. The $N_\mathrm{F}$ dependence of the data is 
compatible with the expectations of random-matrix theory within the large 
statistical errors. Also in this case are needed larger statistical samples 
before drawing any firm conclusion.

\section{Chiral perturbation theory confronts lattice data}
At the NLO in the chiral expansion of the two-flavour theory,
the quark-mass dependence of the pseudoscalar meson mass and 
decay constant is given by
\be\label{eq:NLOlec}
\begin{array}{lcl}
M^2_\pi & = & \displaystyle M^2\, \Big\{1 + \frac{M^2}{32\pi^2F^2} 
            \log\left(\frac{M^2}{\Lambda_3^2}\right) + \dots \Big\}\\[0.5cm]
F_\pi & = & \displaystyle \;\;F\; \Big\{1 - \frac{M^2}{16\pi^2F^2}
          \log\left(\frac{M^2}{\Lambda_4^2}\right) \;+\; \dots \Big\}\; .
\end{array}
\ee
Following the convention 
introduced in Ref.~\cite{Gasser:1983yg}, we can define the 
NLO low-energy constants as
\be
{ \bar l_3} = { \displaystyle
\log\left(\frac{\Lambda^2_3}{M^2}\right)\Big|_{M=139.6 \mathrm{MeV}}}\; , \qquad\qquad
{ \bar l_4} = { \displaystyle
\log\left(\frac{\Lambda^2_4}{M^2}\right)\Big|_{M=139.6 \mathrm{MeV}}} \, .
\ee
They can be determined, at least in principle,
by matching lattice QCD results with the formulas in 
Eqs.~(\ref{eq:NLOlec}). 

\begin{figure}[t]
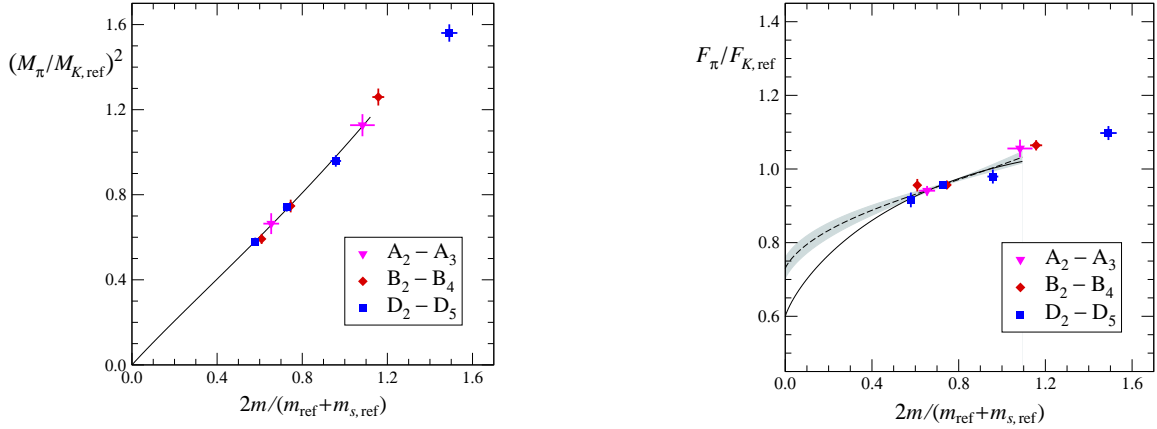

\setlength{\MiniPageLeft}{0.6\textwidth}
\setlength{\MiniPageRight}{\textwidth}\addtolength{\MiniPageRight}{-0.6\textwidth}
\begin{minipage}[t]{\MiniPageLeft}
\vspace{1.0cm}

\includegraphics[width=6.5cm]{mpichpt.eps}
\end{minipage}
\begin{minipage}[t]{\MiniPageRight}
\vspace{1.0cm}

\includegraphics[width=6.125cm]{fpichpt.eps}
\end{minipage}
\caption{The pion mass $M_\pi$ (left) and the pion decay 
constant $F_\pi$ (right) as a function of the bare current quark mass in units of the 
same quantities at the 
reference point from Ref.~\cite{DelDebbio:2006cn}. The solid lines
are fits in the range $M_\pi/M_{K,\mathrm{ref}} \leq 1.1$. using the one-loop 
formulas in Eqs.~(\protect\ref{eq:NLOlecII}). The grey band on the right-hand plot 
is from a fit where a hypothetical next-order term has been included (see text).}\label{fig:ChPT}
\end{figure}

Data available to date do not allow for
a determination of these low-energy constants with a reliable estimate of the errors. 
Nevertheless it is interesting to look in some detail into the analysis presented in  
Ref.~\cite{DelDebbio:2006cn}, which is based on the richest set of data 
at light-quark masses in the two-flavour theory.
The reference point introduced in Section~\ref{sec:rawdata} is a useful idea 
in this context. The dimensionful quantities expressed in units of the 
scales at the reference point are free 
from systematics coming from the determinations of the lattice spacing and 
of ultraviolet renormalization constants. Following 
Ref.~\cite{DelDebbio:2006cn}, we can introduce
\be
 x=\frac{2\,m}{m_\mathrm{ref} + m_{s,\mathrm{ref}}}\; ,\qquad\qquad  
C=\frac{M_{K,\mathrm{ref}}^2}{32 \pi^2 F_{K,\mathrm{ref}}^2},
\ee
and if we define
\be
\hat F = \frac{F}{F_{K,\mathrm{ref}}},\qquad 
\hat B =\frac{m_\mathrm{ref} + m_{s,\mathrm{ref}}}{M_{K,\mathrm{ref}}^2}\,B\,,
\qquad \hat l_n = \log\left(\frac{\Lambda^2_n}{M_{K,\mathrm{ref}}^2}\right)\, ,
\ee
then $\hat l_n=\bar l_n -2.53$, and Eqs.~(\ref{eq:NLOlec}) become
\be\label{eq:NLOlecII}
\begin{array}{lcl}
\displaystyle
\frac{M^2_\pi}{M^2_{K,\mathrm{ref}}} & = & \displaystyle \hat B\, x + 
C\frac{\hat B^2 x^2}{\hat F^2}\left\{\log(\hat B\, x)-\hat l_3\right\}
+ \dots\\[0.5cm]
\displaystyle
\frac{F_\pi}{F_{K,\mathrm{ref}}} & = & \displaystyle \hat F - 
2\,C\, \frac{\hat B\, x}{\hat F}\left\{\log(\hat B\, x)-\hat l_4\right\} 
+\dots
\end{array}
\ee
In the range $M_\pi/M_{K,\mathrm{ref}} \leq 1.1$, the first formula
in Eqs.~(\ref{eq:NLOlecII}) fits the data for the pion mass very well,
the fit parameters being $\hat B =1.11(6)(3)$ and $\hat l_3=0.5(5)(1)$
(equivalently $\bar l_3=3.0(5)(1)$), where the second errors are estimates 
of the systematic uncertainty 
arising from the inaccurately known values of $C$ and $\hat F$; see
Ref.~\cite{DelDebbio:2006cn} for details. The results of the fit 
superimposed on the simulated data is shown in the first plot of 
Fig.~\ref{fig:ChPT}. No attempt has been made to estimate the systematics 
uncertainty coming from higher orders in ChPT, due to discretization effects 
or finite-volume corrections. As described at length in Sections \ref{sec:rawdata} and 
\ref{sec:finvol}, discretization effects seems to be moderate for this quantity, while 
a reliable check of finite-volume effects is still missing and will require simulations
at larger volumes.
A phenomenological analysis of low-energy  experimental 
data gives  $\bar l_3 =  2.9\pm 2.4$~\cite{Gasser:1983yg,Colangelo:2003hf}.
Strictly speaking, the comparison of this value with the one extracted 
from two-flavour lattice simulations only allows for an estimate of the 
contribution to this low-energy constant due to dynamical strange and charm quarks,
assuming that QCD reproduces the experimental data. In practice, however, 
the comparison shows the potentiality of lattice QCD for the 
determination of $\bar l_3$ in the two-flavour theory.

The fit of the data (solid line) for the pseudoscalar decay constant 
with the second of Eqs.~(\ref{eq:NLOlecII}) is shown in the second plot 
of Fig.~\ref{fig:ChPT}. The $\chi^2/\mathrm{d.o.f.}$ turns out 
to be quite good, but the absence of curvature in the data forces the extrapolated
value to be quite lower than the simulated data. A more realistic 
fit (dashed line) can be obtained by including a hypothetical two-loop 
term proportional to $\hat B^2x^2/\hat F^3$ in the second of Eqs.~(\ref{eq:NLOlecII}), 
with a reasonable value of the coefficient $C'=0.046$. The fit parameters $\hat F$ and 
$\hat l_4$ change from $0.60(4)$ and $1.6(1)$ to $0.73(3)$ and $0.73(8)$, respectively, 
when the two-loop term is added~\cite{DelDebbio:2006cn}.  This analysis shows that 
data at significantly 
smaller quark masses, with small systematic and statistical errors, will be required for 
a reliable determination of the parameters in the chiral Lagrangian. 

\section{Conclusions}
Thanks to algorithmic and technical progress achieved over the last couple of years,
it is now possible to simulate QCD with dynamical quarks much more efficiently
than was possible before. As a result lattices with pions as light as 
250--300 MeV can be simulated with the present generation of computers.

Two-flavour QCD is already being simulated with quark masses 
as light as $(m_s/5)$--$(m_s/4)$ with several gluon and fermion 
discretizations, on quite large volumes and fine lattice spacings.
Unquenched effects are clearly seen in various correlation
functions. Maybe to date the most striking result is the observed 
linearity of the pion mass squared as a function of the 
current quark mass in the range $(m_s/4)$--$m_s$. Data at several (fine) 
lattice spacings and with several gluon and fermion actions are consistent with 
this picture. Finite-volume effects are still a concern. A matching 
of the chiral pertubation theory prediction with the finite-size 
effects observed in $M_\pi$ is still missing. In this respect simulations 
at several volumes are still needed. JLQCD started an ambitious project of 
simulating two-flavour QCD at fixed topology with 
exact chiral symmetry.  First preliminary results of these efforts have been already 
presented at this conference.

Several collaborations are already simulating QCD with 
$2+1$ flavours. The RBC and the UKQCD collaborations are attacking the 
problem with domain-wall fermions, while the PACS-CS collaboration is 
working hard to implement a combination of DD-HMC  and PHMC algorithms to 
simulate SW non-perturbative improved fermions.

Thanks to all this work, our community has now the possibility 
to perform many computations which we were dreaming about 
only few years ago. 

\section*{Acknowledegments}

Many thanks to L. Del Debbio, M. L\"uscher, R.~Petronzio and N.~Tantalo
for allowing me to present our results before 
their publication in Refs.~\cite{DelDebbio:2006cn,DelDebbio:2006qab}. 
I warmly thank M.~L\"uscher for many suggestions and endless discussions during 
the preparation of the talk, and for a critical reading of this 
report. It is a pleasure to thank T. Blum, H. Fukaya, S. Hashimoto, 
K.~Jansen, T. Kaneko, C. B. Lang, M. Lin, M. Okamoto and A. Shindler for 
sending their results and for interesting discussions about their work.
An informative e-mail exchange with S. Sharpe about the content of our talks
is acknowledged. Many thanks to the organizers for their great work in 
organizing the conference, and for inviting me to present this talk.
\bibliographystyle{JHEP}
\bibliography{lgiusti_Lat06}

\providecommand{\href}[2]{#2}\begingroup\raggedright\begin{thebibliography}{10}

\bibitem{Aoki:2002fd}
{\bf CP-PACS} Collaboration, S.~Aoki {\em et~al.}, {\it Light hadron spectrum
  and quark masses from quenched lattice {QCD}},  {\em {\rm Phys. Rev.}} {\bf
  D67} (2003) 034503, [\href{http://xxx.lanl.gov/abs/{\rm
  hep-lat/0206009}}{{\tt {\rm hep-lat/0206009}}}].

\bibitem{Eicker:1998sy}
{\bf TXL} Collaboration, N.~Eicker {\em et~al.}, {\it Light and strange hadron
  spectroscopy with dynamical {W}ilson fermions},  {\em {\rm Phys. Rev.}} {\bf
  D59} (1999) 014509, [\href{http://xxx.lanl.gov/abs/{\rm
  hep-lat/9806027}}{{\tt {\rm hep-lat/9806027}}}].

\bibitem{Lippert:1997vg}
T.~Lippert {\em et~al.}, {\it {SESAM} and {T$\chi$L} results for {W}ilson
  action: A status report},  {\em {\rm Nucl. Phys. Proc. Suppl.}} {\bf 60A}
  (1998) 311--334, [\href{http://xxx.lanl.gov/abs/{\rm hep-lat/9707004}}{{\tt
  {\rm hep-lat/9707004}}}].

\bibitem{Allton:2001sk}
{\bf UKQCD} Collaboration, C.~R. Allton {\em et~al.}, {\it Effects of
  non-perturbatively improved dynamical fermions in {QCD} at fixed lattice
  spacing},  {\em {\rm Phys. Rev.}} {\bf D65} (2002) 054502,
  [\href{http://xxx.lanl.gov/abs/{\rm hep-lat/0107021}}{{\tt {\rm
  hep-lat/0107021}}}].

\bibitem{AliKhan:2001tx}
{\bf CP-PACS} Collaboration, A.~Ali~Khan {\em et~al.}, {\it Light hadron
  spectroscopy with two flavors of dynamical quarks on the lattice},  {\em {\rm
  Phys. Rev.}} {\bf D65} (2002) 054505, {Erratum--{\it ibid.} {\bf D67} (2003)
  059901}, [\href{http://xxx.lanl.gov/abs/{\rm hep-lat/0105015}}{{\tt {\rm
  hep-lat/0105015}}}].

\bibitem{Aoki:2002uc}
{\bf JLQCD} Collaboration, S.~Aoki {\em et~al.}, {\it Light hadron spectroscopy
  with two flavors of {${\cal O}(a)$}-improved dynamical quarks},  {\em {\rm
  Phys. Rev.}} {\bf D68} (2003) 054502, [\href{http://xxx.lanl.gov/abs/{\rm
  hep-lat/0212039}}{{\tt {\rm hep-lat/0212039}}}].

\bibitem{Duane:1987de}
S.~Duane, A.~D. Kennedy, B.~J. Pendleton, and D.~Roweth, {\it Hybrid {M}onte
  {C}arlo},  {\em {\rm Phys. Lett.}} {\bf B195} (1987) 216--222.

\bibitem{Bernard:2002pd}
C.~Bernard {\em et~al.}, {\it Panel discussion on the cost of dynamical quark
  simulations},  {\em {\rm Nucl. Phys. Proc. Suppl.}} {\bf 106} (2002)
  199--205.

\bibitem{Bernard:2001av}
C.~W. Bernard {\em et~al.}, {\it The {QCD} spectrum with three quark flavors},
  {\em {\rm Phys. Rev.}} {\bf D64} (2001) 054506,
  [\href{http://xxx.lanl.gov/abs/{\rm hep-lat/0104002}}{{\tt {\rm
  hep-lat/0104002}}}].

\bibitem{Sharpe:2006re}
S.~R. Sharpe, {\it Rooted staggered fermions: good, bad or ugly?},
  \href{http://xxx.lanl.gov/abs/{\rm hep-lat/0610094}}{{\tt {\rm
  hep-lat/0610094}}}.

\bibitem{Luscher:2003vf}
M.~L{\"u}scher, {\it Lattice {QCD} and the {S}chwarz alternating procedure},
  {\em {\rm JHEP}} {\bf 05} (2003) 052, [\href{http://xxx.lanl.gov/abs/{\rm
  hep-lat/0304007}}{{\tt {\rm hep-lat/0304007}}}].

\bibitem{Luscher:2003qa}
M.~L{\"u}scher, {\it Solution of the {D}irac equation in lattice {QCD} using a
  domain decomposition method},  {\em {\rm Comput. Phys. Commun.}} {\bf 156}
  (2004) 209--220, [\href{http://xxx.lanl.gov/abs/{\rm hep-lat/0310048}}{{\tt
  {\rm hep-lat/0310048}}}].

\bibitem{Luscher:2004rx}
M.~L{\"u}scher, {\it Schwarz-preconditioned {HMC} algorithm for two-flavour
  lattice {QCD}},  {\em {\rm Comput. Phys. Commun.}} {\bf 165} (2005) 199--220,
  [\href{http://xxx.lanl.gov/abs/{\rm hep-lat/0409106}}{{\tt {\rm
  hep-lat/0409106}}}].

\bibitem{Hasenbusch:2001ne}
M.~Hasenbusch, {\it Speeding up the hybrid-{M}onte-{C}arlo algorithm for
  dynamical fermions},  {\em {\rm Phys. Lett.}} {\bf B519} (2001) 177--182,
  [\href{http://xxx.lanl.gov/abs/{\rm hep-lat/0107019}}{{\tt {\rm
  hep-lat/0107019}}}].

\bibitem{Urbach:2005ji}
C.~Urbach, K.~Jansen, A.~Shindler, and U.~Wenger, {\it {HMC} algorithm with
  multiple time scale integration and mass preconditioning},  {\em {\rm Comput.
  Phys. Commun.}} {\bf 174} (2006) 87--98, [\href{http://xxx.lanl.gov/abs/{\rm
  hep-lat/0506011}}{{\tt {\rm hep-lat/0506011}}}].

\bibitem{DelDebbio:2005qa}
L.~Del~Debbio, L.~Giusti, M.~L{\"u}scher, R.~Petronzio, and N.~Tantalo, {\it
  Stability of lattice {QCD} simulations and the thermodynamic limit},  {\em
  {\rm JHEP}} {\bf 02} (2006) 011, [\href{http://xxx.lanl.gov/abs/{\rm
  hep-lat/0512021}}{{\tt {\rm hep-lat/0512021}}}].

\bibitem{DelDebbio:2006cn}
L.~Del~Debbio, L.~Giusti, M.~L{\"u}scher, R.~Petronzio, and N.~Tantalo, {\it
  {QCD} with light {W}ilson quarks on fine lattices ({I}): first experiences
  and physics results},  \href{http://xxx.lanl.gov/abs/{\rm
  hep-lat/0610059}}{{\tt {\rm hep-lat/0610059}}}.

\bibitem{DelDebbio:2006qab}
L.~Del~Debbio, L.~Giusti, M.~L{\"u}scher, R.~Petronzio, and N.~Tantalo, {\it
  {QCD} with light {W}ilson quarks on fine lattices ({II}): {DD-HMC}
  simulations and data analysis},  {\em {\rm to appear}}.

\bibitem{Sheikholeslami:1985ij}
B.~Sheikholeslami and R.~Wohlert, {\it Improved continuum limit lattice action
  for {QCD} with {W}ilson fermions},  {\em {\rm Nucl. Phys.}} {\bf B259} (1985)
  572.

\bibitem{Jansen:1998mx}
{\bf ALPHA} Collaboration, K.~Jansen and R.~Sommer, {\it {${\cal O}(\alpha)$}
  improvement of lattice {QCD} with two flavors of {W}ilson quarks},  {\em {\rm
  Nucl. Phys.}} {\bf B530} (1998) 185--203, Erratum--{\it ibid} {\bf B643}
  (2002) 517--518, [\href{http://xxx.lanl.gov/abs/{\rm hep-lat/9803017}}{{\tt
  {\rm hep-lat/9803017}}}].

\bibitem{Ukawa:2002pc}
{\bf CP-PACS and JLQCD} Collaboration, A.~Ukawa, {\it Computational cost of
  full {QCD} simulations experienced by {CP-PACS} and {JLQCD} collaborations},
  {\em {\rm Nucl. Phys. Proc. Suppl.}} {\bf 106} (2002) 195--196.

\bibitem{Sexton:1992nu}
J.~C. Sexton and D.~H. Weingarten, {\it Hamiltonian evolution for the hybrid
  {M}onte {C}arlo algorithm},  {\em {\rm Nucl. Phys.}} {\bf B380} (1992)
  665--678.

\bibitem{Gockeler:2006ns}
M.~G{\"o}ckeler {\em et~al.}, {\it Simulating at realistic quark masses:
  pseudoscalar decay constants and chiral logarithms},
  \href{http://xxx.lanl.gov/abs/{\rm hep-lat/0610066}}{{\tt {\rm
  hep-lat/0610066}}}.

\bibitem{Jansen:2006rf}
{\bf ETM} Collaboration, K.~Jansen and C.~Urbach, {\it First results with two
  light flavours of quarks with maximally twisted mass},
  \href{http://xxx.lanl.gov/abs/{\rm hep-lat/0610015}}{{\tt {\rm
  hep-lat/0610015}}}.

\bibitem{Clark:2006fx}
M.~A. Clark and A.~D. Kennedy, {\it Accelerating dynamical fermion computations
  using the rational hybrid {M}onte {C}arlo ({RHMC}) algorithm with multiple
  pseudofermion fields},  \href{http://xxx.lanl.gov/abs/{\rm
  hep-lat/0608015}}{{\tt {\rm hep-lat/0608015}}}.

\bibitem{Clark:2006wq}
M.~A. Clark, {\it The rational hybrid {M}onte {C}arlo algorithm},
  \href{http://xxx.lanl.gov/abs/{\rm hep-lat/0610048}}{{\tt {\rm
  hep-lat/0610048}}}.

\bibitem{Niedermayer:1998bi}
F.~Niedermayer, {\it Exact chiral symmetry, topological charge and related
  topics},  {\em {\rm Nucl. Phys. Proc. Suppl.}} {\bf 73} (1999) 105--119,
  [\href{http://xxx.lanl.gov/abs/{\rm hep-lat/9810026}}{{\tt {\rm
  hep-lat/9810026}}}].

\bibitem{Wilke:1997gf}
T.~Wilke, T.~Guhr, and T.~Wettig, {\it The microscopic spectrum of the {QCD}
  {D}irac operator with finite quark masses},  {\em {\rm Phys. Rev.}} {\bf D57}
  (1998) 6486--6495, [\href{http://xxx.lanl.gov/abs/{\rm hep-th/9711057}}{{\tt
  {\rm hep-th/9711057}}}].

\bibitem{Giusti:2003gf}
L.~Giusti, M.~L{\"u}scher, P.~Weisz, and H.~Wittig, {\it Lattice {QCD} in the
  epsilon-regime and random matrix theory},  {\em {\rm JHEP}} {\bf 11} (2003)
  023, [\href{http://xxx.lanl.gov/abs/{\rm hep-lat/0309189}}{{\tt {\rm
  hep-lat/0309189}}}].

\bibitem{Fukaya:2006xp}
{\bf JLQCD} Collaboration, H.~Fukaya {\em et~al.}, {\it Dynamical overlap
  fermions in the epsilon-regime},  \href{http://xxx.lanl.gov/abs/{\rm
  hep-lat/0610024}}{{\tt {\rm hep-lat/0610024}}}.

\bibitem{Aoki:1983qi}
S.~Aoki, {\it New phase structure for lattice {QCD} with {W}ilson fermions},
  {\em {\rm Phys. Rev.}} {\bf D30} (1984) 2653.

\bibitem{Frezzotti:2000nk}
{\bf ALPHA} Collaboration, R.~Frezzotti, P.~A. Grassi, S.~Sint, and P.~Weisz,
  {\it Lattice {QCD} with a chirally twisted mass term},  {\em {\rm JHEP}} {\bf
  08} (2001) 058, [\href{http://xxx.lanl.gov/abs/{\rm hep-lat/0101001}}{{\tt
  {\rm hep-lat/0101001}}}].

\bibitem{Sharpe:2006ia}
S.~R. Sharpe, {\it Discretization errors in the spectrum of the {H}ermitian
  {W}ilson-{D}irac operator},  {\em {\rm Phys. Rev.}} {\bf D74} (2006) 014512,
  [\href{http://xxx.lanl.gov/abs/{\rm hep-lat/0606002}}{{\tt {\rm
  hep-lat/0606002}}}].

\bibitem{Luscher:1992an}
M.~L{\"u}scher, R.~Narayanan, P.~Weisz, and U.~Wolff, {\it The
  {S}chr{\"o}dinger functional: A renormalizable probe for non abelian gauge
  theories},  {\em {\rm Nucl. Phys.}} {\bf B384} (1992) 168--228,
  [\href{http://xxx.lanl.gov/abs/{\rm hep-lat/9207009}}{{\tt {\rm
  hep-lat/9207009}}}].

\bibitem{Luscher:1991wu}
M.~L{\"u}scher, P.~Weisz, and U.~Wolff, {\it A numerical method to compute the
  running coupling in asymptotically free theories},  {\em {\rm Nucl. Phys.}}
  {\bf B359} (1991) 221--243.

\bibitem{Jansen:1995ck}
K.~Jansen {\em et~al.}, {\it Non-perturbative renormalization of lattice {QCD}
  at all scales},  {\em {\rm Phys. Lett.}} {\bf B372} (1996) 275--282,
  [\href{http://xxx.lanl.gov/abs/{\rm hep-lat/9512009}}{{\tt {\rm
  hep-lat/9512009}}}].

\bibitem{Luscher:1998pe}
M.~L{\"u}scher, {\it Advanced lattice {QCD}, in: {P}robing the {S}tandard
  {M}odel of {P}article {I}nteractions ({L}es {H}ouches 1997), {E}ds. {R}.
  {G}upta et al. ({E}lsevier, {A}msterdam, 1999)},
  \href{http://xxx.lanl.gov/abs/{\rm hep-lat/9802029}}{{\tt {\rm
  hep-lat/9802029}}}.

\bibitem{Luscher:1992zx}
M.~L{\"u}scher, R.~Sommer, U.~Wolff, and P.~Weisz, {\it Computation of the
  running coupling in the {SU(2)} {Y}ang-{M}ills theory},  {\em {\rm Nucl.
  Phys.}} {\bf B389} (1993) 247--264, [\href{http://xxx.lanl.gov/abs/{\rm
  hep-lat/9207010}}{{\tt {\rm hep-lat/9207010}}}].

\bibitem{Luscher:1993gh}
M.~L{\"u}scher, R.~Sommer, P.~Weisz, and U.~Wolff, {\it A precise determination
  of the running coupling in the {SU(3)} {Y}ang-{M}ills theory},  {\em {\rm
  Nucl. Phys.}} {\bf B413} (1994) 481--502, [\href{http://xxx.lanl.gov/abs/{\rm
  hep-lat/9309005}}{{\tt {\rm hep-lat/9309005}}}].

\bibitem{DellaMorte:2004bc}
{\bf ALPHA} Collaboration, M.~Della~Morte {\em et~al.}, {\it Computation of the
  strong coupling in {QCD} with two dynamical flavours},  {\em {\rm Nucl.
  Phys.}} {\bf B713} (2005) 378--406, [\href{http://xxx.lanl.gov/abs/{\rm
  hep-lat/0411025}}{{\tt {\rm hep-lat/0411025}}}].

\bibitem{Sommer:1993ce}
R.~Sommer, {\it A new way to set the energy scale in lattice gauge theories and
  its applications to the static force and {$alpha_s$} in {SU(2)}
  {Y}ang-{M}ills theory},  {\em {\rm Nucl. Phys.}} {\bf B411} (1994) 839--854,
  [\href{http://xxx.lanl.gov/abs/{\rm hep-lat/9310022}}{{\tt {\rm
  hep-lat/9310022}}}].

\bibitem{Gockeler:2004rp}
{\bf QCDSF} Collaboration, M.~G{\"o}ckeler {\em et~al.}, {\it Determination of
  light and strange quark masses from full lattice {QCD}},  {\em {\rm Phys.
  Lett.}} {\bf B639} (2006) 307--311, [\href{http://xxx.lanl.gov/abs/{\rm
  hep-ph/0409312}}{{\tt {\rm hep-ph/0409312}}}].

\bibitem{Meyer:2006ty}
H.~B. Meyer {\em et~al.}, {\it Exploring the {HMC} trajectory-length dependence
  of autocorrelation times in lattice {QCD}},
  \href{http://xxx.lanl.gov/abs/{\rm hep-lat/0606004}}{{\tt {\rm
  hep-lat/0606004}}}.

\bibitem{DellaMorte:2005kg}
{\bf ALPHA} Collaboration, M.~Della~Morte {\em et~al.}, {\it Non-perturbative
  quark mass renormalization in two-flavor {QCD}},  {\em {\rm Nucl. Phys.}}
  {\bf B729} (2005) 117--134, [\href{http://xxx.lanl.gov/abs/{\rm
  hep-lat/0507035}}{{\tt {\rm hep-lat/0507035}}}].

\bibitem{DellaMorte:2005rd}
M.~Della~Morte, R.~Hoffmann, F.~Knechtli, R.~Sommer, and U.~Wolff, {\it
  Non-perturbative renormalization of the axial current with dynamical {W}ilson
  fermions},  {\em {\rm JHEP}} {\bf 07} (2005) 007,
  [\href{http://xxx.lanl.gov/abs/{\rm hep-lat/0505026}}{{\tt {\rm
  hep-lat/0505026}}}].

\bibitem{Allton:2004qq}
{\bf UKQCD} Collaboration, C.~R. Allton {\em et~al.}, {\it Improved {W}ilson
  {QCD} simulations with light quark masses},  {\em {\rm Phys. Rev.}} {\bf D70}
  (2004) 014501, [\href{http://xxx.lanl.gov/abs/{\rm hep-lat/0403007}}{{\tt
  {\rm hep-lat/0403007}}}].

\bibitem{Aoki:2004ht}
Y.~Aoki {\em et~al.}, {\it Lattice {QCD} with two dynamical flavors of domain
  wall fermions},  {\em {\rm Phys. Rev.}} {\bf D72} (2005) 114505,
  [\href{http://xxx.lanl.gov/abs/{\rm hep-lat/0411006}}{{\tt {\rm
  hep-lat/0411006}}}].

\bibitem{Mawhinney:2006lat}
{\bf RBC and UKQCD} Collaboration, R.~Mawhinney {\em et~al.}, {\it Production
  and properties of $2+1$ flavor {DWF} ensembles},  {\em {\rm these
  proceedings}}.

\bibitem{Maynard:2006lat}
{\bf RBC and UKQCD} Collaboration, C.~Maynard {\em et~al.}, {\it Baryon
  spectrum in $2+1$ flavour domain wall {QCD} from {QCDOC}},  {\em {\rm these
  proceedings}}.

\bibitem{Lin:2006lat}
{\bf RBC and UKQCD} Collaboration, M.~Lin {\em et~al.}, {\it Chiral
  extrapolations in $2+1$ flavor domain wall fernions simulations},  {\em {\rm
  these proceedings}}.

\bibitem{Allton:2006ax}
{\bf RBC and UKQCD} Collaboration, C.~Allton {\em et~al.}, {\it Light meson
  masses and non-perturbative renormalisation in 2+1 flavour domain wall qcd},
  \href{http://xxx.lanl.gov/abs/{\rm hep-lat/0610119}}{{\tt {\rm
  hep-lat/0610119}}}.

\bibitem{Ukawa:2006pre}
{\bf PACS-CS} Collaboration, A.~Ukawa {\em et~al.}, {\it Status and physics
  plan of the {PACS-CS} project},  {\em {\rm PoS}} {\bf LAT2006} (2006) 039.

\bibitem{Ishikawa:2006pb}
{\bf PACS-CS} Collaboration, K.-I. Ishikawa {\em et~al.}, {\it An application
  of the {UV}-filtering preconditioner to the polynomial hybrid {M}onte {C}arlo
  algorithm},  \href{http://xxx.lanl.gov/abs/{\rm hep-lat/0610037}}{{\tt {\rm
  hep-lat/0610037}}}.

\bibitem{Kuramashi:2006np}
{\bf PACS-CS} Collaboration, Y.~Kuramashi {\em et~al.}, {\it 2+1 flavor lattice
  {QCD} with {L}{\"u}scher's domain-decomposed {HMC} algorithm},
  \href{http://xxx.lanl.gov/abs/{\rm hep-lat/0610063}}{{\tt {\rm
  hep-lat/0610063}}}.

\bibitem{Allton:1996yv}
C.~R. Allton, V.~Gimenez, L.~Giusti, and F.~Rapuano, {\it Light quenched hadron
  spectrum and decay constants on different lattices},  {\em {\rm Nucl. Phys.}}
  {\bf B489} (1997) 427--452, [\href{http://xxx.lanl.gov/abs/{\rm
  hep-lat/9611021}}{{\tt {\rm hep-lat/9611021}}}].

\bibitem{Heitger:2000ay}
{\bf ALPHA} Collaboration, J.~Heitger, R.~Sommer, and H.~Wittig, {\it Effective
  chiral lagrangians and lattice {QCD}},  {\em {\rm Nucl. Phys.}} {\bf B588}
  (2000) 377--399, [\href{http://xxx.lanl.gov/abs/{\rm hep-lat/0006026}}{{\tt
  {\rm hep-lat/0006026}}}].

\bibitem{Luscher:2005mv}
M.~L{\"u}scher, {\it Lattice {QCD} with light {W}ilson quarks},  {\em {\rm
  PoS}} {\bf LAT2005} (2006) 002, [\href{http://xxx.lanl.gov/abs/{\rm
  hep-lat/0509152}}{{\tt {\rm hep-lat/0509152}}}].

\bibitem{Sharpe:1998xm}
S.~R. Sharpe and R.~L. Singleton, {\it Spontaneous flavor and parity breaking
  with {W}ilson fermions},  {\em {\rm Phys. Rev.}} {\bf D58} (1998) 074501,
  [\href{http://xxx.lanl.gov/abs/{\rm hep-lat/9804028}}{{\tt {\rm
  hep-lat/9804028}}}].

\bibitem{Gockeler:2006vi}
M.~G{\"o}ckeler {\em et~al.}, {\it Simulating at realistic quark masses: light
  quark masses},  \href{http://xxx.lanl.gov/abs/{\rm hep-lat/0610071}}{{\tt
  {\rm hep-lat/0610071}}}.

\bibitem{Gasser:1986vb}
J.~Gasser and H.~Leutwyler, {\it Light quarks at low temperatures},  {\em {\rm
  Phys. Lett.}} {\bf B184} (1987) 83.

\bibitem{Gasser:1987ah}
J.~Gasser and H.~Leutwyler, {\it Thermodynamics of chiral symmetry},  {\em {\rm
  Phys. Lett.}} {\bf B188} (1987) 477.

\bibitem{Gasser:1987zq}
J.~Gasser and H.~Leutwyler, {\it Spontaneously broken symmetries: effective
  lagrangians at finite volume},  {\em {\rm Nucl. Phys.}} {\bf B307} (1988)
  763.

\bibitem{Fukugita:1992jj}
M.~Fukugita, H.~Mino, M.~Okawa, G.~Parisi, and A.~Ukawa, {\it Finite size
  effect for hadron masses in lattice {QCD}},  {\em {\rm Phys. Lett.}} {\bf
  B294} (1992) 380--384.

\bibitem{Orth:2005kq}
B.~Orth, T.~Lippert, and K.~Schilling, {\it Finite-size effects in lattice
  {QCD} with dynamical {W}ilson fermions},  {\em {\rm Phys. Rev.}} {\bf D72}
  (2005) 014503, [\href{http://xxx.lanl.gov/abs/{\rm hep-lat/0503016}}{{\tt
  {\rm hep-lat/0503016}}}].

\bibitem{Hashimoto:2006rb}
{\bf JLQCD} Collaboration, S.~Hashimoto {\em et~al.}, {\it Dynamical overlap
  fermion at fixed topology},  \href{http://xxx.lanl.gov/abs/{\rm
  hep-lat/0610011}}{{\tt {\rm hep-lat/0610011}}}.

\bibitem{Matsufuru:2006xr}
{\bf JLQCD} Collaboration, H.~Matsufuru {\em et~al.}, {\it Improvement of
  algorithms for dynamical overlap fermions},
  \href{http://xxx.lanl.gov/abs/{\rm hep-lat/0610026}}{{\tt {\rm
  hep-lat/0610026}}}.

\bibitem{Kaneko:2006pa}
{\bf JLQCD} Collaboration, T.~Kaneko {\em et~al.}, {\it {JLQCD}'s dynamical
  overlap project},  \href{http://xxx.lanl.gov/abs/{\rm hep-lat/0610036}}{{\tt
  {\rm hep-lat/0610036}}}.

\bibitem{Fukaya:2006vs}
{\bf JLQCD} Collaboration, H.~Fukaya {\em et~al.}, {\it Lattice gauge action
  suppressing near-zero modes of {$H(W)$}},  \href{http://xxx.lanl.gov/abs/{\rm
  hep-lat/0607020}}{{\tt {\rm hep-lat/0607020}}}.

\bibitem{Neuberger:1997fp}
H.~Neuberger, {\it Exactly massless quarks on the lattice},  {\em {\rm Phys.
  Lett.}} {\bf B417} (1998) 141--144, [\href{http://xxx.lanl.gov/abs/{\rm
  hep-lat/9707022}}{{\tt {\rm hep-lat/9707022}}}].

\bibitem{Gasser:1983yg}
J.~Gasser and H.~Leutwyler, {\it Chiral perturbation theory to one loop},  {\em
  {\rm Ann. Phys.}} {\bf 158} (1984) 142.

\bibitem{Colangelo:2003hf}
G.~Colangelo and S.~Durr, {\it The pion mass in finite volume},  {\em {\rm Eur.
  Phys. J.}} {\bf C33} (2004) 543--553, [\href{http://xxx.lanl.gov/abs/{\rm
  hep-lat/0311023}}{{\tt {\rm hep-lat/0311023}}}].

\end{thebibliography}\endgroup
\end{document}